\pdfoutput=1

\documentclass[11pt]{article}

\usepackage[]{acl}

\usepackage{times}
\usepackage{latexsym}
\usepackage{enumitem}
\usepackage[T1]{fontenc}

\usepackage[utf8]{inputenc}

\usepackage{microtype}

\usepackage{algorithm}
\usepackage{algorithmic}
\usepackage{amsmath}
\usepackage{amssymb}
\usepackage{multirow}
\usepackage{caption}
\usepackage{subcaption}
\usepackage[most]{tcolorbox}
\usepackage{color, colortbl}
\definecolor{Gray}{gray}{0.9}
\usepackage{todonotes}
\small{
} 

\title{The Inefficiency of Language Models in Scholarly Retrieval: An Experimental Walk-through}

\author{Shruti Singh
  \and
  Mayank Singh \\
  Department of Computer Science and Engineering \\
  Indian Institute of Technology Gandhinagar \\
  Gujarat, India \\
  \texttt{\{singh\_shruti, singh.mayank\}@iitgn.ac.in} \\}

\begin{document}
\maketitle
\begin{abstract}
Language models are increasingly becoming popular in AI-powered scientific IR systems. This paper evaluates popular scientific language models in handling (i) \textit{short-query texts} and (ii)  \textit{textual neighbors}. Our experiments showcase the inability to retrieve relevant documents for a short-query text even under the most relaxed conditions.  Additionally, we leverage textual neighbors, generated by small perturbations to the original text, to demonstrate that not all perturbations lead to close neighbors in the embedding space. Further, an exhaustive categorization yields several classes of orthographically and semantically related, partially related and completely unrelated neighbors.  
Retrieval performance turns out to be more influenced by the surface form rather than the semantics of the text.
\end{abstract}

\section{Introduction}
Representation learning methods have drastically evolved large scientific volume exploration strategies. The popular applications include summarization, construction of mentor-mentee network~\citep{ke2021dataset}, recommendation~\citep{ostendorff2020aspect,specter2020cohan,das2020information,hope-etal-2021-extracting}, QA over scientific documents~\citep{DBLP:conf/emnlp/SuXYSBF20}, and verification of scientific claims~\citep{Wadden2020FactOF}. The growing community's interest has led to the development of several scientific document embedding models such as OAG-BERT~\citep{liu2021oag}, SPECTER~\citep{specter2020cohan}, SciBERT~\citep{beltagy-etal-2019-scibert}, and BioBERT~\citep{lee2020biobert} over the past five years. 
OAG-BERT has been deployed in the Aminer production system. Given similar possibilities of future deployments of scientific document embeddings models in the existing scholarly systems, it is crucial to evaluate and identify limitations robustly. However, we do not find any existing work that critically analyzes the scientific language models to the best of our knowledge. 

To motivate the reader, we present a simple experiment. Queries \textit{`document vector'} and \textit{`document vectors'} fetch no common candidates among first page results on Google Scholar and Semantic Scholar (candidates in Appendix~\ref{sec:GSSSCandidates}). This illustrates the extremely brittle nature of such systems to minor alterations in query text, leading to completely different search outcomes.
To motivate further, we experiment with textual queries encoded by the popular SciBERT model.
A perturbed text \textit{`documen vector'} (relevant in AI) is closer to the Biomedical term \textit{`Virus vector'} in the embedding space. Similar observations were found for many other queries. As scholarly search and recommendation systems are complex systems and their detailed algorithms are not publicly available, we analyze the behavior of scientific language models, which are (or will be) presumably an integral component of each of these systems. Motivated by the usage of perturbed inputs to stress test ML systems in interpretability analysis, we propose to use \textit{`textual neighbors'} to analyze how they are represented in the embedding space of scientific LMs. Unlike previous works~\citep{ribeiro-etal-2020-beyond,rychalska2019models}, which analyze the effect of perturbations on downstream task-specific models, we focus on analyzing the embeddings which are originally inputs to such downstream models. With the explosion of perturbation techniques for various kinds of robustness and interpretability analysis, it is difficult to generalize the insights gathered from perturbation experiments. We propose a classification schema based on orthography and semantics, to organize the perturbation strategies.

The distribution of various types of textual neighbors in a training corpus is non-uniform. Specifically, the low frequency of textual neighbors results in non-optimized representations, wherein semantically similar neighbors might end up distant in the space. These non-optimal representations have a cascading effect on the downstream task-specific models. In this work, we analyze whether all textual neighbors of input \textit{X} are also \textit{X}’s neighbors in the embedding space. Further, we also study if their presence in the embedding space can negatively impact downstream applications dependent on similarity-based document retrieval. Our main contributions are:
\begin{enumerate}[noitemsep,nolistsep]
\item We introduce (in Section~\ref{sec:tncategories}) five textual neighbor categories based on orthography and semantics. We further construct a non-exhaustive list of thirty-two textual neighbor types and organize them into these five categories to analyze the behavior of scientific LMs on manipulated text.
\item We conduct (in Section~\ref{sec:short-text}) robust experiments to showcase the limitations of scientific LMs under a short-text query retrieval scheme. 
\item We analyze (in Section~\ref{sec:textual_neighbors}) embeddings of textual neighbors and their placement in the embedding space of three scientific LMs, namely SciBERT, SPECTER, and OAG-BERT. Our experiments highlight the capability and limitations of different models in representing different categories of textual neighbors.
\end{enumerate}

\section{Related Works}
Several works utilize Textual Neighbors to interpret decisions of classifiers~\citep{lime,gardner-etal-2020-evaluating}, test linguistic capabilities of NLP models~\citep{ribeiro-etal-2020-beyond}, measure progress in language generation~\citep{DBLP:journals/corr/abs-2102-01672}, and generate semantically equivalent adversarial examples~\citep{ribeiro-etal-2018-semantically}.
Similar to these works, we use textual neighbors of scientific papers to analyze the behavior of scientific LMs.~\citet{macavaney2020abnirml} analyze the behavior of neural IR models by proposing test strategies: constructing test samples by controlling specific measures (e.g., term frequency, document length), and by manipulating text (e.g., removing stops, shuffling words). This is closest to our work, as we also employ text manipulation to analyze the behavior of scientific language models using a simple Alternative-Self Retrieval scheme (Section~\ref{sec:exp_design}). However, our focus is not evaluation of retrieval-augmented models and we only use a relaxed document retrieval pipeline in our evaluation to analyze the behavior of scientific LMs trained on diverse domains, in encoding scientific documents. We organize the textual neighbors into categories which capture different capabilities of LMs. We also show that it is crucial to evaluate models on dissimilar texts rather than just semantically similar textual neighbors. Ours is a first work in analyzing the properties of scientific LMs for different inputs and can be utilized by future works to design and evaluate future scientific LMs. Due to space limitations, we present the detailed discussion on scientific LMs in Appendix~\ref{sec:app_scilm}.

\section{Short Queries and Textual Neighbors}
\label{sec:tncategories}
In this paper, we experiment with short queries to fetch relevant scientific documents. The term \textit{`short'} signifies a query length comparable to the length of research titles. The candidates are constructed from either title (T) or title and abstract (T+A) text. We, further, make small alterations to the candidate text to construct \textit{`textual neighbors'}. The textual neighbors can be syntactically, semantically, or structurally similar to the candidate text. Unlike previous works that explore textual neighbors to analyze and stress test complex models (Q\&A, Sentiment, NLI, NER~\citep{rychalska2019models}), we experiment directly with representation learning models and analyze the placement of textual neighbors in their embedding space. 
While semantically similar neighbors are frequently used in previous works~\citep{ribeiro-etal-2018-semantically}; we also explore \textit{semantically dissimilar} textual neighbors to analyze scientific language models. While an LM is expected to represent semantically similar texts with high similarity, some orthographically similar but semantically dissimilar texts can have highly similar embeddings, which is undesirable behavior.  Note that we restrict the current query set to titles for two main reasons: (i) most real-world search queries are shorter in length, and (ii) flat keyword-based search lacks intent and can lead to erroneous conclusions.  

\begin{table}[!t]
\centering
\resizebox{\columnwidth}{!}{
\small{
\begin{tabular}{lp{0.13\columnwidth}l}
\rowcolor{Gray}
\multicolumn{1}{l}{\textbf{O-S}} &
\multicolumn{1}{l}{\textbf{Neighbor Code}} & \multicolumn{1}{l}{\textbf{Form}} \\ \hline
LL-PS & T\_ARot & \begin{tabular}[c]{@{}l@{}}Title $\rightarrow$ Preserve\\ Abs $\rightarrow$ Rotate\end{tabular} \\ \hline
LL-PS & T\_AShuff & \begin{tabular}[c]{@{}l@{}}Title $\rightarrow$ Preserve\\ Abs $\rightarrow$ Shuffle\end{tabular} \\ \hline
LL-PS & T\_ASortAsc & \begin{tabular}[c]{@{}l@{}}Title $\rightarrow$ Preserve\\ Abs $\rightarrow$ Sort Ascending\end{tabular} \\ \hline
LL-PS & T\_ASortDesc & \begin{tabular}[c]{@{}l@{}}Title $\rightarrow$ Preserve\\ Abs $\rightarrow$ Sort Descending\end{tabular} \\ \hline
LO-PS & T\_ADelRand & \begin{tabular}[c]{@{}l@{}}Title $\rightarrow$ Preserve\\ Abs $\rightarrow$ Random word deletion 30\%\end{tabular} \\ \hline
LO-PS & T\_ADelADJ & \begin{tabular}[c]{@{}l@{}}Title $\rightarrow$ Preserve\\ Abs $\rightarrow$ Delete all ADJs\end{tabular} \\ \hline
LO-DS & T\_ADelNN & \begin{tabular}[c]{@{}l@{}}Title $\rightarrow$ Preserve\\ Abs $\rightarrow$ Delete all NNs\end{tabular} \\ \hline
LO-PS & T\_ADelVB & \begin{tabular}[c]{@{}l@{}}Title $\rightarrow$ Preserve\\ Abs $\rightarrow$ Delete all Verbs\end{tabular} \\ \hline
LO-PS & T\_ADelADV & \begin{tabular}[c]{@{}l@{}}Title $\rightarrow$ Preserve\\ Abs $\rightarrow$ Delete all ADVs\end{tabular} \\ \hline
LO-PS & T\_ADelPR & \begin{tabular}[c]{@{}l@{}}Title $\rightarrow$ Preserve\\ Abs $\rightarrow$ Delete all PRs\end{tabular} \\ \hline
LO-HS & T\_ADelDT & \begin{tabular}[c]{@{}l@{}}Title $\rightarrow$ Preserve\\ Abs $\rightarrow$ Delete all DTs\end{tabular} \\ \hline
LO-PS & T\_ADelNum & \begin{tabular}[c]{@{}l@{}}Title $\rightarrow$ Preserve\\ Abs $\rightarrow$ Delete all Numbers\end{tabular} \\ \hline
LO-DS & T\_ADelNNPH & \begin{tabular}[c]{@{}l@{}}Title $\rightarrow$ Preserve\\ Abs $\rightarrow$ Delete  all NN Phrases\end{tabular} \\ \hline
LO-PS & T\_ADelTopNNPH & \begin{tabular}[c]{@{}l@{}}Title $\rightarrow$ Preserve\\ Abs $\rightarrow$ Delete top 50\% NPs\end{tabular} \\ \hline
LO-HS & TDelADJ\_A & \begin{tabular}[c]{@{}l@{}}Title $\rightarrow$ Delete all ADJs\\ Abs $\rightarrow$ Preserve\end{tabular} \\ \hline
LO-HS & TDelNN\_A & \begin{tabular}[c]{@{}l@{}}Title $\rightarrow$ Delete all NNs\\ Abs $\rightarrow$ Preserve\end{tabular} \\ \hline
LO-HS & TDelVB\_A & \begin{tabular}[c]{@{}l@{}}Title $\rightarrow$ Delete all VBs\\ Abs $\rightarrow$ Preserve\end{tabular} \\ \hline
LO-HS & TDelDT\_A & \begin{tabular}[c]{@{}l@{}}Title $\rightarrow$ Delete all DTs\\ Abs $\rightarrow$ Preserve\end{tabular} \\ \hline
LO-DS & TDelNN & \begin{tabular}[c]{@{}l@{}}Title $\rightarrow$ Delete all NNs\\ Abs $\rightarrow$ Delete\end{tabular} \\ \hline
LO-PS & T\_ADelQ1 & \begin{tabular}[c]{@{}l@{}}Title $\rightarrow$ Preserve\\ Abs $\rightarrow$ Delete quantile 1\end{tabular} \\ \hline
LO-PS & T\_ADelQ2 & \begin{tabular}[c]{@{}l@{}}Title $\rightarrow$ Preserve\\ Abs $\rightarrow$ Delete quantile 2\end{tabular} \\ \hline
LO-PS & T\_ADelQ3 & \begin{tabular}[c]{@{}l@{}}Title $\rightarrow$ Preserve\\ Abs $\rightarrow$ Delete quantile 3\end{tabular} \\ \hline
LL-HS & TNNU\_A & \begin{tabular}[c]{@{}l@{}}Title $\rightarrow$ Uppercase NNs\\ Abs $\rightarrow$ Preserve\end{tabular} \\ \hline
LL-HS & TNonNNU\_A & \begin{tabular}[c]{@{}l@{}}Title $\rightarrow$ Uppercase non NNs\\ Abs $\rightarrow$ Preserve\end{tabular} \\ \hline
LL-HS & T\_ANNU & \begin{tabular}[c]{@{}l@{}}Title $\rightarrow$ Preserve\\ Abs $\rightarrow$ Uppercase NNs\end{tabular} \\ \hline
LL-HS & T\_ANonNNU & \begin{tabular}[c]{@{}l@{}}Title $\rightarrow$ Preserve\\ Abs $\rightarrow$ Uppercase non NNs\end{tabular} \\ \hline
LO-PS & T\_A\_DelNNChar & \begin{tabular}[c]{@{}l@{}}Title $\rightarrow$ Delete chars from NNs\\ Abs $\rightarrow$ Delete chars from NNs\end{tabular} \\ \hline
LO-HS & TRepNNT\_A & \begin{tabular}[c]{@{}l@{}}Title $\rightarrow$ Add a NN from title\\ Abs $\rightarrow$ Preserve\end{tabular} \\ \hline
LO-HS & TRepNNA\_A & \begin{tabular}[c]{@{}l@{}}Title $\rightarrow$ Add a NN from abs\\ Abs $\rightarrow$ Preserve\end{tabular} \\ \hline
LO-DS & T\_ADelNonNNs & \begin{tabular}[c]{@{}l@{}}Title $\rightarrow$ Preserve\\ Abs $\rightarrow$ Delete all non NNs\end{tabular} \\ \hline
LO-DS & T\_ARepADJ & \begin{tabular}[c]{@{}l@{}}Title $\rightarrow$ Preserve\\ Abs $\rightarrow$ Replace ADJs with antonyms\end{tabular} \\ \hline
LL-HS & T\_A\_WS & \begin{tabular}[c]{@{}l@{}}Randomly replace 50\% whitespace \\ chars with 2-5 whitespace chars\\ in the title \& abstract\end{tabular} \\ \hline
\end{tabular}
}}
\caption{Neighbor code is in format Txx\_Ayy\_zz, where xx and yy denote the perturbation to the paper title (T) and the abstract (A) respectively. zz denotes perturbation to both T and A. Missing T or A denotes that the corresponding input field is deleted completely.}
\label{tab:tneighbors}
\end{table}

Textual neighbors have a similar word form or similar meaning. 
We observe that Textual neighbors possess the following properties based on two aspects: (i) Orthography (surface-level information content of text in terms of characters, words, and sentences), and (ii) Semantics. These two aspects are integral for a pair of texts to be textual neighbors. The properties of these aspects are:
\begin{enumerate}
    \item \textbf{Orthography}: Orthographic-neighbors are generated by making small surface-level transformations to the input. Based on the textual content, neighbors can be generated by: 
    \begin{enumerate}[noitemsep,nolistsep]
        \item \textbf{\textit{Lossy Perturbation (LO)}}: `Lossy' behavior indicates that the information from the original text is lost, either due to addition or deletion of textual content. E.g., random deletion of 2--3 characters from a word.
        \item \textbf{\textit{Lossless Perturbation (LL)}}: `Lossless' behavior indicates that the original information content is unaltered. E.g., shuffling sentences of a paragraph.
    \end{enumerate}
    \item \textbf{Semantics}: Textual neighbors can either be semantically similar or not. While semantic categories are subjective, we try to define them objectively: 
    \begin{enumerate}[noitemsep,nolistsep]
        \item \textbf{\textit{Dissimilar (DS)}}: Semantically dissimilar neighbors are generated by deletion of nouns (NNs) or noun phrases (NPs), or by changing the directionality of text such as replacing adjectives (ADJs) with antonyms.
        \item \textbf{\textit{Partially Similar (PS)}}: Partially similar neighbors are constructed by sentence level modifications such as sentences scrambling (while preserving the word order in each sentence), word deletion of non-NNs (or NPs), or character deletion in words (including NNs)  such that the textual neighbor preserves at least 70\% of the words in the original text.
        \item \textbf{\textit{Highly Similar (HS)}}: Highly similar neighbors are constructed by (i) non-word-semantic changes such as changing the case of words or altering the whitespace characters, (ii) word deletions or additions such that the textual neighbor preserves at least 90\% of the words in the original text. Note that deletion of NNs only from the title while preserving the abstract will qualify to be the \textit{HS} category.
    \end{enumerate}
\end{enumerate}

Since SPECTER and OAG-BERT utilize paper titles and abstracts to learn embeddings, we generate textual neighbors by altering texts of these two input fields. We present 32 textual neighbor types in Table~\ref{tab:tneighbors}. Each of these 32 neighbor types is categorized into one of the five categories: LO-HS, LO-PS, LO-DS, LL-HS, and LL-PS. We exclude the LL-DS category (e.g., scrambling all words in the text) as it is infrequent and less probable to occur in a real-world setting. Examples of the textual neighbors are presented in Appendix~\ref{sec:app_tneigh} (Table~\ref{apptab:examples_tneig}).

\begin{table}[!tb]
\newcolumntype{L}{>{\centering\arraybackslash}p{0.11\linewidth}}
\newcolumntype{K}{>{\centering\arraybackslash}p{0.38\linewidth}}
\centering
\resizebox{\columnwidth}{!}{
\small{
\begin{tabular}{lLKr}
\hline
\rowcolor{Gray}
\textbf{Dataset} & \textbf{Domain} & \textbf{SubDomain} & \textbf{\#Papers} \\ \hline
\textbf{ACL-ANTH} & CS & NLP & 35,482 \\ \hline
\textbf{ICLR} & CS & Deep Learning & 5,032 \\ \hline
\textbf{Arxiv-CS-SY} & CS & Systems and Control & 10,000\\ \hline
\textbf{Arxiv-MATH} & MT & Algebraic Topology & 10,000\\ \hline
\textbf{Arxiv-HEP} & HEP & HEP Theory & 10,000\\ \hline
\textbf{Arxiv-QBIO} & QB & Neurons and Cognition & 6,903 \\ \hline
\textbf{Arxiv-ECON} & ECO &  Econometrics, General \& Theoretical Economics & 4,542 \\ \hline
\end{tabular}}
}
\caption{Dataset Statistics}
\label{tab:dataset}
\end{table}

\begin{figure}[!t]
    \centering
    \begin{tabular}{l}
    \includegraphics[width=1.0\columnwidth]{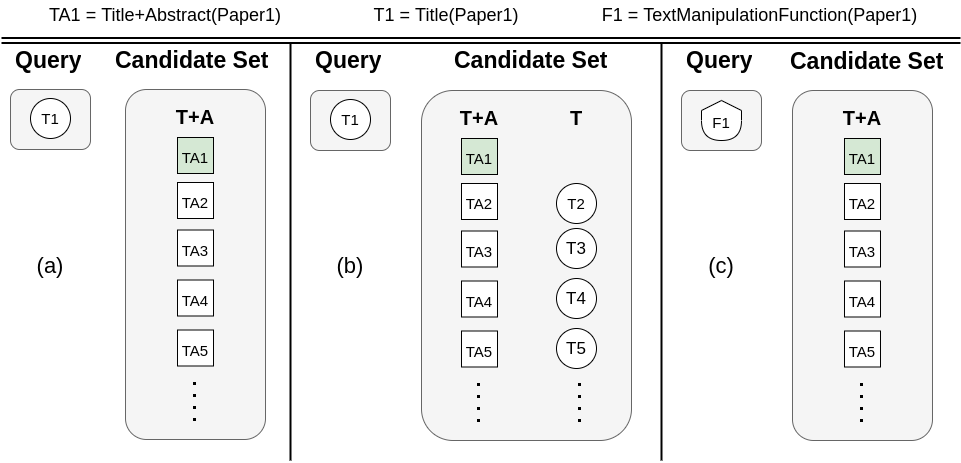} 
    \end{tabular}
    \caption{Alternative-Self Retrieval schemes for (a) Sec.~\ref{sec:short-text} Task I, (b) Sec.~\ref{sec:short-text} Task II, and (c) Sec.~\ref{ssec:altret_tneigh}. Green represents the relevant candidate document for the query. The query is a subset of the relevant candidate document in schemes (a) and (b), and a textual neighbor of the relevant candidate in scheme (c).}
    \label{fig:altret_scheme}
\end{figure}

\section{Experiment Design}
\label{sec:exp_design}
\noindent \textbf{The Alternative-Self Retrieval:} We propose an embarrassingly simple binary retrieval scheme which contains only one relevant document in the candidate set. Alternative-Self Retrieval refers to the characteristic that the query is an altered version of the relevant candidate document. E.g., the candidate documents are the embeddings of paper title and abstract (henceforth T+A) and the query is embedding of title. We present a schematic of three Alternative-Self Retrieval schemes in Figure~\ref{fig:altret_scheme}. The retrieval is simple and we measure performance with accommodating metrics discussed in this section further. Our purpose is to analyze scientific LM embeddings under the most relaxed conditions.

\noindent \textbf{The Datasets:} We evaluate the scientific LMs on seven datasets (statistics in Table~\ref{tab:dataset}) to understand their effectiveness in encoding documents from diverse research fields. Each dataset contains the titles and abstracts of papers. We curate the ACL Anthology dataset\footnote{https://aclanthology.org/} and the ICLR dataset from OpenReview\footnote{https://openreview.net/group?id=ICLR.cc}.
To control the size of the ACL Anthology dataset, we exclude papers from workshops and non-ACL venues. 
We also curate five datasets from arXiv for the domains Mathematics (MT), High Energy Physics (HEP), Quantitative Biology (QB), Economics (ECO), and Computer Science (CS). We make available our code and dataset for public access\footnote{https://github.com/shruti-singh/scilm\_exp}.

\noindent \textbf{The Notations:}
\textbf{$\mathcal{D}$} is the set of seven datasets described in Table~\ref{tab:dataset}.  \textbf{$\mathcal{X}$} is the set of original input texts to the scientific LMs consisting of the paper title (T) and the abstract (A). For d $\in \mathcal{D}$, $\mathcal{X}$ = \{$x_j: x_j = \mbox{(T+A)}(p)$, where $\mbox{(T+A)}(p)=$ concat(title($p$), abstract($p$), $\forall \mbox{ paper }p \in$ d\}. $f$  represents the type of textual neighbor (represented by the neighbor code presented in Table~\ref{tab:tneighbors}). $\mathcal{Q}$ and $\mathcal{R}$ are the query and candidate set for the IR task.

\noindent \textbf{Evaluation Metrics:} We report performance scores on the following retrieval metrics: \newline
\underline{\emph{Mean Reciprocal Rank (MRR)}}: All our tasks use binary relevance of documents to compute MRR. \newline
\underline{\emph{T100}}: It represents the percentage of queries which retrieve the one and only relevant document among the top-100 documents. \newline
\underline{\emph{NN\textit{k}\_Ret}}: \% of queries in textual neighbor category whose k nearest neighbors (k-NN) retrieve the original document. \newline
\underline{\emph{AOP-10}}: Average overlap percentage among 10-NN of x and y, where x = T+A($x_j$) and y = $f(x_j)$. $f$ is a text manipulation function, represented by the textual neighbor codes presented in Table~\ref{tab:tneighbors}. \newline

\begin{table*}[!tb]
\centering
\small{
\begin{tabular}{l||cc|cc|cc||cc|cc|cc}
\hline
& \multicolumn{6}{c||}{\textbf{Task I:}  $\mathcal{R} = \mathcal{X}$} & \multicolumn{6}{c}{\textbf{Task II:}  $\mathcal{R} = \mathcal{X}$ $\cup$ T($\mathcal{X}$)} \\ \hline
& \multicolumn{2}{c|}{\textbf{SciBERT}} & \multicolumn{2}{c|}{\textbf{SPECTER}} & \multicolumn{2}{c||}{\textbf{OAG-BERT}} & \multicolumn{2}{c|}{\textbf{SciBERT}} & \multicolumn{2}{c|}{\textbf{SPECTER}} & \multicolumn{2}{c}{\textbf{OAG-BERT}} \\ \hline
& \textit{\textbf{MRR}} & \textit{\textbf{T100}} & \textit{\textbf{MRR}} & \textit{\textbf{T100}} & \textit{\textbf{MRR}} & \textit{\textbf{T100}} & \textit{\textbf{MRR}} & \textit{\textbf{T100}} & \textit{\textbf{MRR}} & \textit{\textbf{T100}} & \textit{\textbf{MRR}} & \textit{\textbf{T100}} \\ \hline
\rowcolor{Gray}
\textbf{Arxiv-MATH} & 0.007 & 5.7 & \textbf{0.596} & \textbf{90.5} & 0.143 & 25.1 & 0 & 0 & \textbf{0.276} & \textbf{64.2} & 0.133 & 31.2 \\ \hline

\textbf{Arxiv-HEP} & 0.006 & 5.8 & \textbf{0.693} & \textbf{92.4} & 0.182 & 29.7 & 0 & 0 & \textbf{0.354} & \textbf{75.7} & 0.193 & 39.0 \\ \hline

\rowcolor{Gray}
\textbf{Arxiv-QBIO} & 0.008 & 6.6 & \textbf{0.789} & \textbf{97.5} & 0.187 & 33.0  & 0 & 0.2 & \textbf{0.507} & \textbf{90.6} & 0.194 & 36.1 \\ \hline

\textbf{Arxiv-ECON} & 0.009 & 10.4 & \textbf{0.783} & \textbf{96.2} & 0.177 & 32.5 & 0 & 0.1 & \textbf{0.519} & \textbf{87.5} & 0.176 & 37.2 \\ \hline

\rowcolor{Gray}
\textbf{Arxiv-CS\_SY} & 0.011 & 5.0 & \textbf{0.859} & \textbf{99.2} & 0.186 & 32.1 & 0 & 0.2 & \textbf{0.553} & \textbf{92.9} & 0.163 & 34.0 \\ \hline

\textbf{ICLR} & 0.004 & 5.5 & \textbf{0.586} & \textbf{91.5} & 0.140 & 29.8 & 0 & 0 & \textbf{0.221} & \textbf{66.5} & 0.128 & 33.0 \\ \hline

\rowcolor{Gray}
\textbf{ACL-ANTH} & 0.002 & 2.3 & \textbf{0.739} & \textbf{94.3} &  0.138 & 26.5 & 0 & 0 & \textbf{0.315} & \textbf{77.0} & 0.101 & 27.3 \\ \hline
\end{tabular}}
\caption{For both Task I and Task II, SPECTER consistently performs the best on all datasets. For Task II, drop in MRR and T100 scores for SPECTER is significant in comparison to Task I.}
\label{tab:exp1exp2}
\end{table*}

\begin{figure*}[!thb]
     \centering
     \begin{subfigure}[b]{0.32\textwidth}
         \centering
         \includegraphics[width=\textwidth]{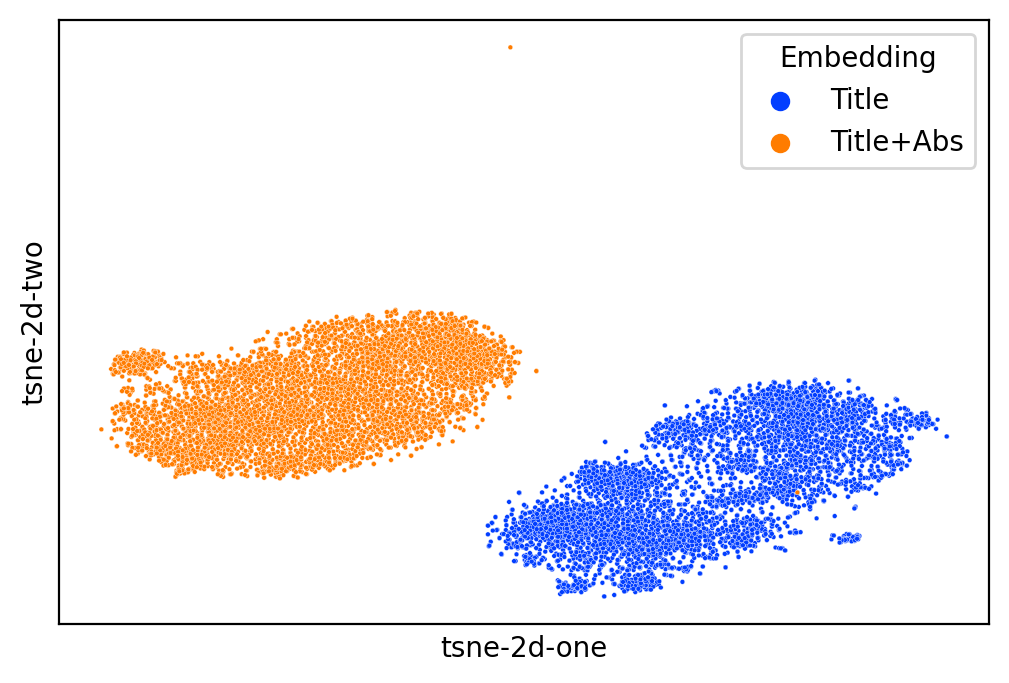}
         \caption{SciBERT}
     \end{subfigure}
     \hfill
     \begin{subfigure}[b]{0.32\textwidth}
         \centering
         \includegraphics[width=\textwidth]{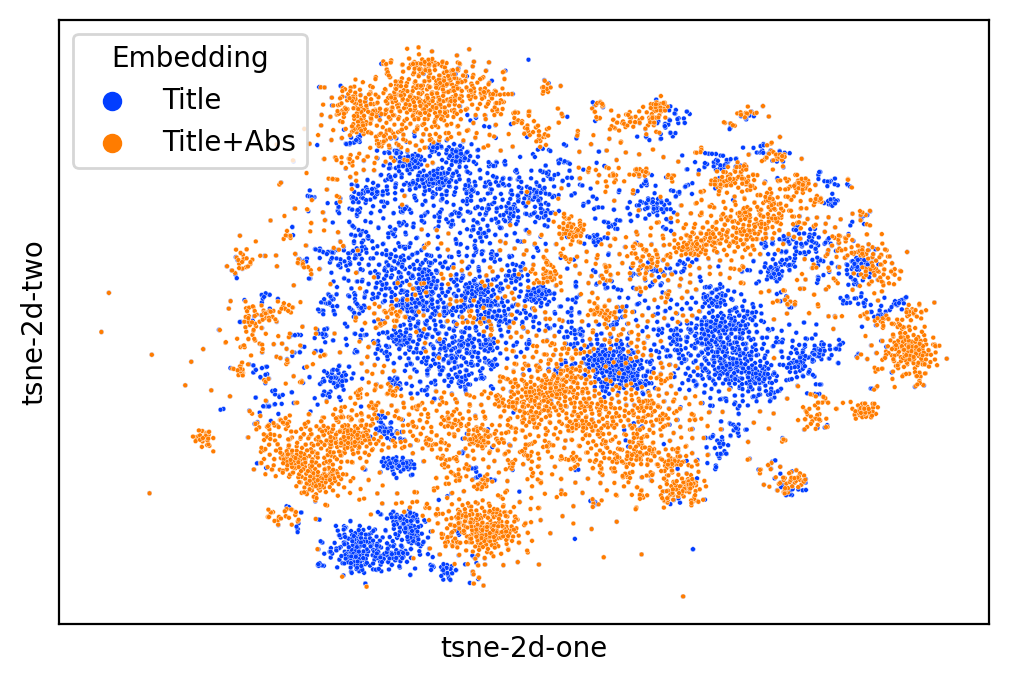}
         \caption{SPECTER}
     \end{subfigure}
     \hfill
     \begin{subfigure}[b]{0.32\textwidth}
         \centering
         \includegraphics[width=\textwidth]{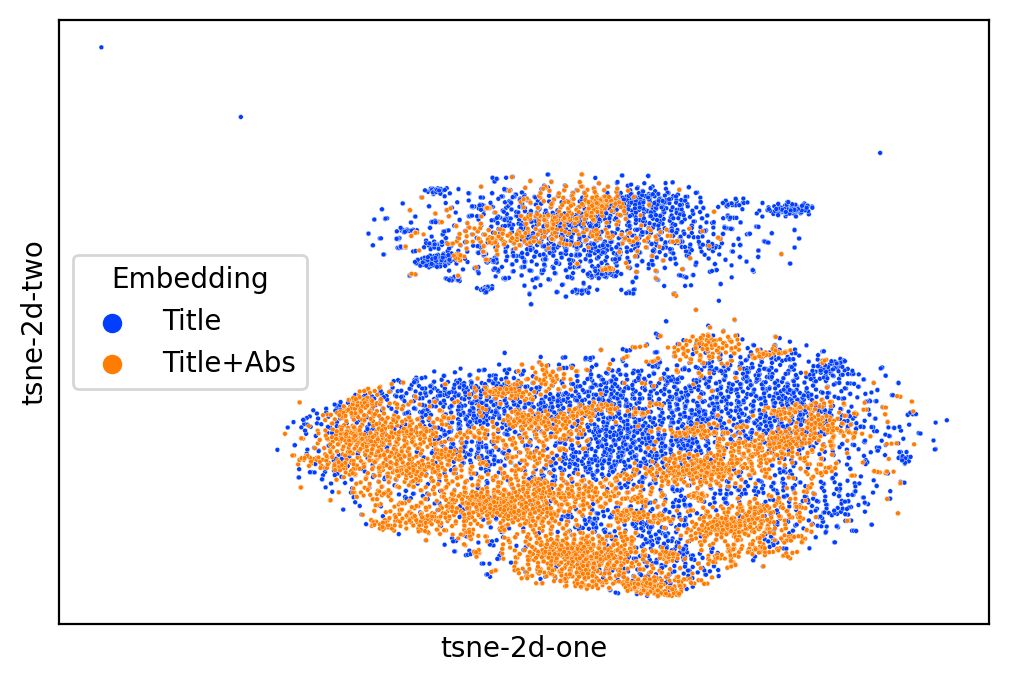}
         \caption{OAG-BERT}
     \end{subfigure}
    \caption{t-SNE plots for T and T+A embeddings for the ICLR dataset. Completely non-overlapping embeddings for T and T+A by SciBERT model highlight differences in encoding texts of varying lengths.}
    \label{fig:e2_tsne}
\end{figure*}

\begin{figure}[!t]
     \centering
     \begin{subfigure}[b]{0.48\columnwidth}
         \centering
         \includegraphics[width=\textwidth]{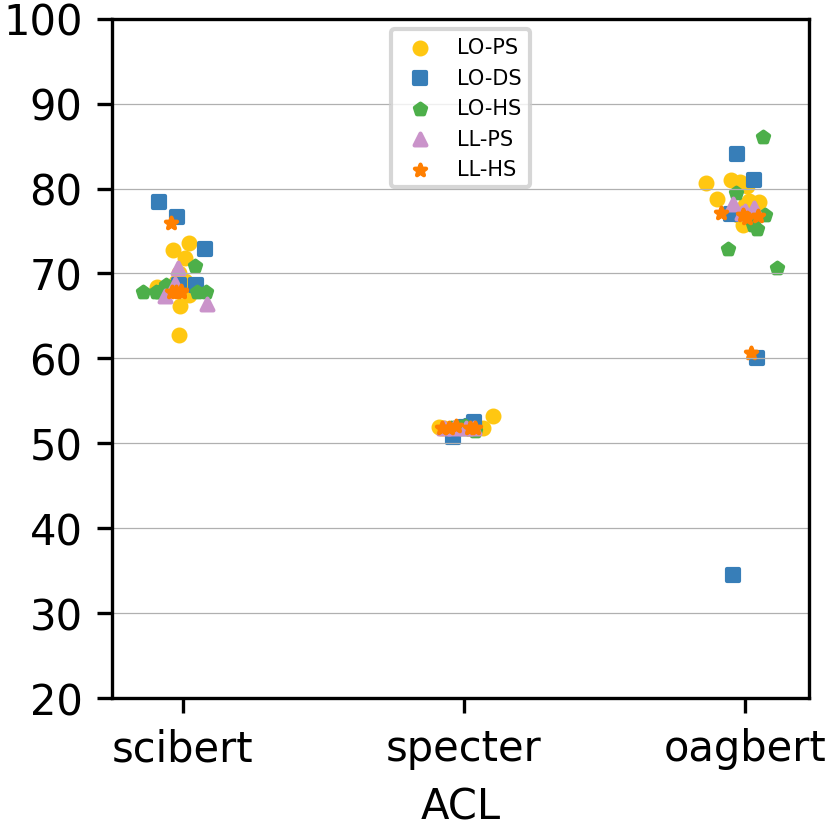}
     \end{subfigure}
     \hfill
     \begin{subfigure}[b]{0.48\columnwidth}
         \centering
         \includegraphics[width=\textwidth]{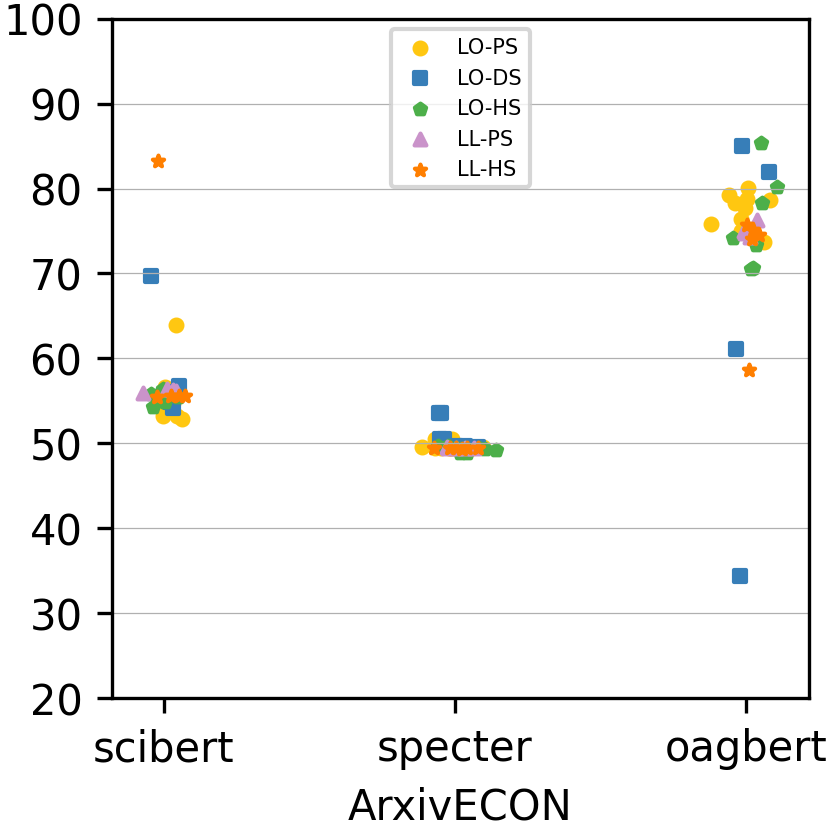}
     \end{subfigure}
    \caption{Percentage of pair of documents for each Textual Neighbor whose similarity is greater than the average similarity. OAG-BERT has high inter similarity ($>$ 50\%), i.e. more than 50\% document pairs have cosine similarity greater than average similarity.}
    \label{fig:per_hs}
\end{figure}

\begin{figure}[!tb]
    \centering
    \begin{subfigure}[b]{0.95\columnwidth}
        \centering
        \includegraphics[trim={0.1cm 0 0.1cm 0},clip,width=\textwidth]{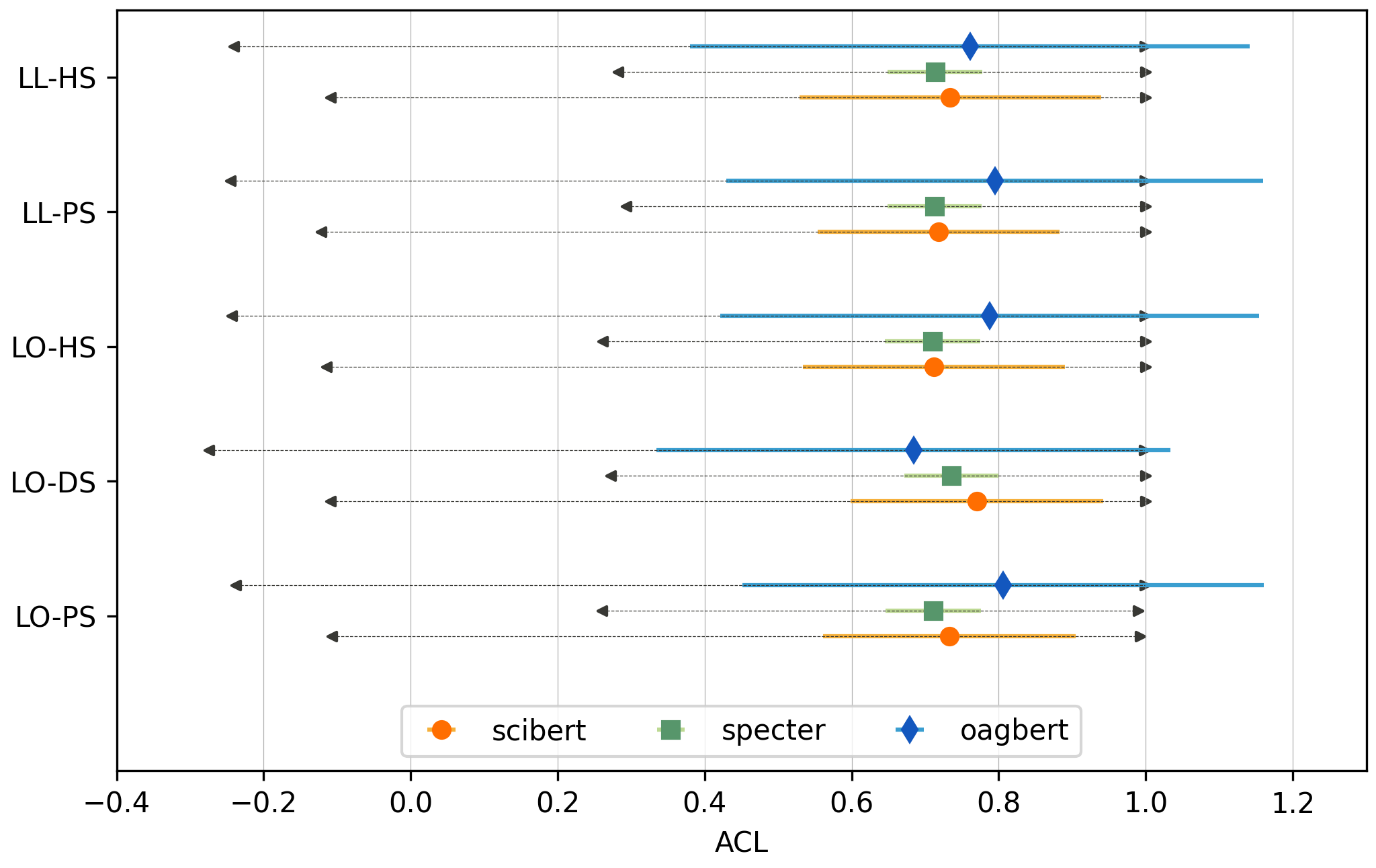}
    \end{subfigure}
    \caption{Inter Similarity of Textual Neighbor Vectors. Bold lines represent the $\mu$ and the $\sigma$ of pairwise similarities. Arrow heads represent min and max values. Pairwise similarities are spread out in a broad range for OAG-BERT, suggesting vectors for textual neighbors are more spread out in the vector space.
    }
    \label{fig:geo}
\end{figure}

\section{Analysing Embeddings for Scientific Document Titles and Abstracts}
\label{sec:short-text}
In this section, we experiment with the inputs to the scientific language models. Due to free availability and ease of parsing paper title and abstract, majority scientific LMs learn embeddings from the title and the abstract of the paper. However, multiple downstream applications such as document search involve short queries (often keywords). We present two Alternative-Self retrieval experiments to compare the embeddings of paper title (T) with the embeddings of paper titles and abstract (T+A). In both experiments, $|\mathcal{Q}|$ = 1000 queries for each dataset.
\newline \newline
\textbf{Task I: Querying titles against original candidate documents}
\label{sec:exp1}
In this Alternative-Self Retrieval setup, given a query q constructed only from the paper title, the system recommends relevant candidate document embeddings constructed from title and abstract both (T+A). This setting is similar to querying in a scientific literature search engine as the search queries are usually short. The motivation behind this experiment is to analyse the similarity among the T and the T+A embedding of a paper (Figure~\ref{fig:altret_scheme}(a)).
The experiment details are:\newline
\textit{\textbf{Query}}: $\mathcal{Q}$ = \{$q_j$: $x_j \in \mathcal{X}, f$= T, $q_j$=$f(x_j)$\}\newline
\textit{\textbf{Candidate Documents}}: $\mathcal{R}$ = \{$x_k$: $x_k \in \mathcal{X}$\}\newline
\textit{\textbf{Task}}: For $q_j \in \mathcal{Q}$, rank the candidates based on cosine similarity.\newline
\textit{\textbf{Evaluation}}: MRR and T100. For each $q_j$, there is only one relevant document in the candidate set which is the corresponding T+A embedding.\newline
We present the results for various models on different domains in Table~\ref{tab:exp1exp2}. The results suggest that SciBERT performs poorly for all domains. OAG-BERT on an average ranks the original document in the range 5-7th position. However, we also observe that even in the best case, only 33\% queries retrieve the original document in the top-100 retrieved candidates. SPECTER, on the other hand performs consistently better than both SciBERT and OAG-BERT. The MRR score suggests that on average, the original document is ranked in the top-2 documents, also has a good T100 score across all domains. However, for around 10\% of the queries in the Arxiv-MATH, Arxiv-HEP, and ICLR datasets, SPECTER does not rank the original relevant document in the top-100 retrieved candidates.
\newline
\textbf{Task II: Introducing all Titles in the Candidate Set}
\label{sec:exp2}
To increase the complexity of the previous task, we add all the title embeddings (T) in the candidate set (Figure~\ref{fig:altret_scheme}(b)). We test if the T embeddings are more similar to other titles, or to their corresponding T+A embeddings. \\
\textit{\textbf{Query}}: $\mathcal{Q}$ = \{$q_j$: $x_j \in \mathcal{X}$, $f=$ T, $q_j$=$f(x_j)$\}\newline
\textit{\textbf{Candidate Documents}}: For query $q_j$, the candidate set $\mathcal{R}_j$ is defined as, \\
$\mathcal{R}_j$=\{$f(x_i): x_i\in \mathcal{X}, f=\mbox{T, }i\ne j$\} $\bigcup$ \{$x_k$: $x_k \in \mathcal{X}$\} \newline
The task and evaluation metrics are same as Task I.
Extremely poor values for SciBERT (Table~\ref{tab:exp1exp2}) lead us to examine the vector space of embeddings presented in Figure~\ref{fig:e2_tsne} (t-SNE plots for T and T+A embeddings), revealing that T and T+A embeddings form two non overlapping clusters. Even though the title text is a subset of T+A, SciBERT embeddings are significantly different, suggesting that input length influences SciBERT. This highlights the issue in retrieval for varying length query and candidates.
We present the t-SNE plots for other datasets in the Appendix~\ref{sec:app_embscidoc}. 

SPECTER still performs the best, but a significant drop in MRR and T100  suggests that both T and T+A embeddings tightly cluster together in partially overlapping small groups (can be verified from Figure~\ref{fig:e2_tsne}). However, comparable T100 for OAG-BERT to the previous experiment suggests that the model does not falter when the input text length is small. Ideally, we expect T and T+A embeddings to overlap, indicating that the embeddings of the same paper are closer. The pretraining of these models could be the reason for such distribution of T and T+A embeddings. SciBERT is trained on sentences from full text of research papers, leading to different representations for short (T) and longer (T+A) texts. As SPECTER and OAG-BERT are trained on title and abstract fields both, such non-overlapping behavior is not observed. 

\section{Analysing Scientific LMs with Textual Neighbors}
\label{sec:textual_neighbors}
In the previous section, we experimented with different input fields (T vs T+A). In this section, we experiment with the 32 textual neighbors classes (which alter different input fields: T, A, or T+A). We present our results for the following experiments for the five broad categories: LL-HS, LL-PS, LL-DS, LO-HS, and LO-DS. Due to space constraints, we present plots for selective datasets for each of the experiment in the paper. The rest of the plots are presented in Appendix~\ref{sec:app_scilmtneigh}.

\subsection{Distribution of Textual Neighbors in the Embedding Space}
We measure how textual neighbor embeddings are distributed in the embedding space in each dataset when encoded by the SciBERT, SPECTER, and OAG-BERT model. For each textual neighbor class listed in Table~\ref{tab:tneighbors}, we compute pairwise similarities among all input pairs. A plot of the similarity values for different textual neighbor categories is presented in Figure~\ref{fig:geo} (additional plots in Appendix~\ref{sec:app_simtneighorg}). It can be observed that the pairwise similarities among documents are spread out in a significantly broader range for OAG-BERT than SciBERT and SPECTER on all datasets. The average similarity for all datasets by all models is above 0.5. We do not observe any significant difference in the average similarity for different textual neighbor classes. Interestingly, for the SPECTER model, the minimum similarity is greater than zero for all datasets across all neighbor categories. Document pair similarity via OAG-BERT embeddings have a low average similarity for the LO-DS category. 

We present the percentage of pair of documents for each Textual Neighbor whose similarity is greater than the average similarity in Figure~\ref{fig:per_hs} (additional plots in Appendix~\ref{sec:app_simtneighorg}). OAG-BERT shows high inter similarity (greater than 50\%) for majority of textual neighbors suggesting that more than 50\% document pairs have a cosine similarity greater than average similarity. For SPECTER vectors, all types of textual neighbors have around 50\% documents pairs having a similarity greater than mean similarity. However, extremely high values of percentage of document pairs having similarity greater than average similarity for the SciBERT model on the ACL dataset, and the OAG-BERT on almost all dataset suggests that majority of the documents in the embedding space are represented compactly for all textual neighbor categories. \\

\begin{figure}[!t]
    \centering
    \begin{subfigure}[b]{0.48\columnwidth}
        \centering
        \includegraphics[trim={0.2cm 0.2cm 0.1cm 0},clip,width=\textwidth]{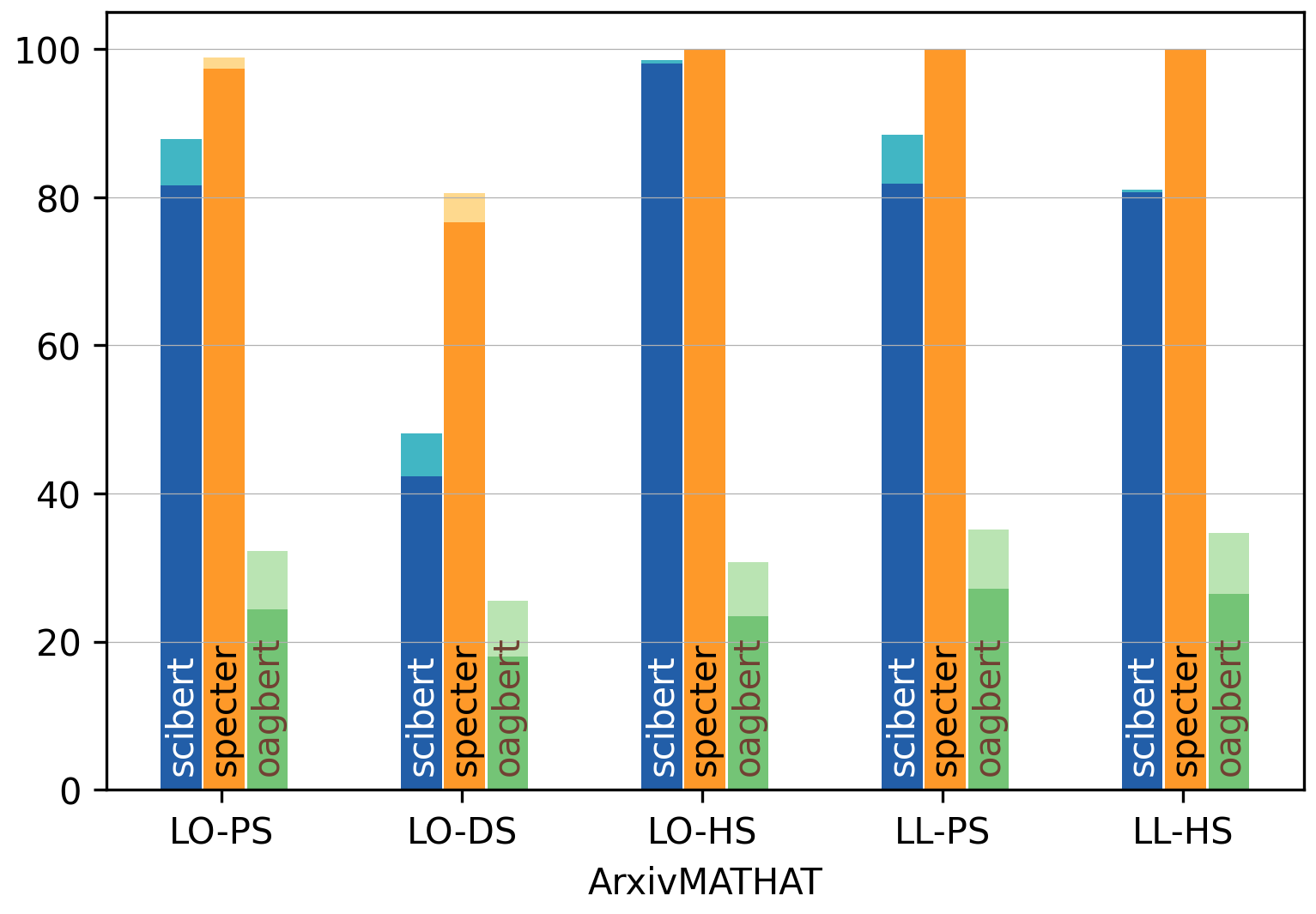}
    \end{subfigure}
    \hfill
    \begin{subfigure}[b]{0.48\columnwidth}
        \centering
        \includegraphics[trim={0.2cm 0.2cm 0.1cm 0},clip,width=\textwidth]{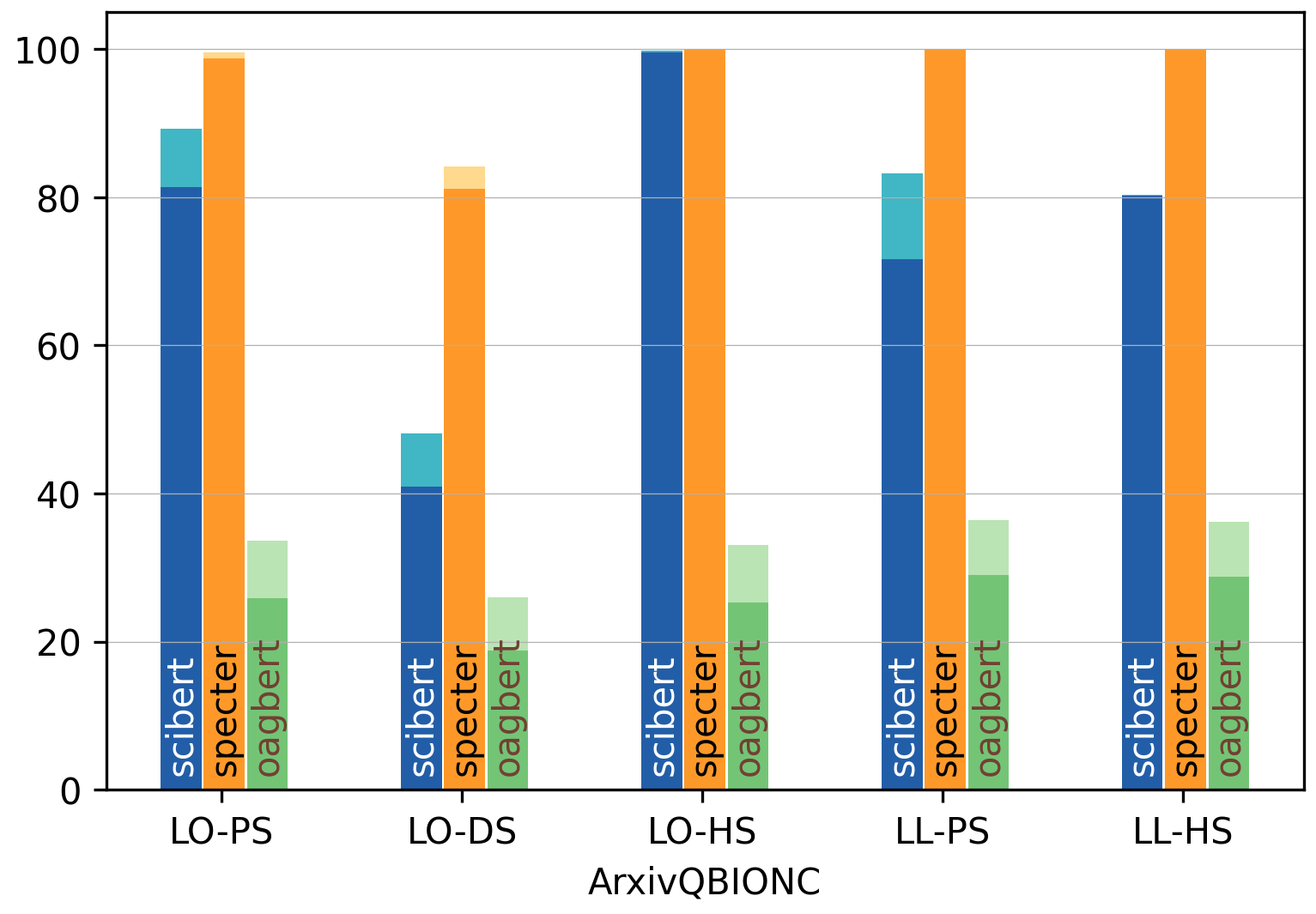}
    \end{subfigure}
    \caption{The bottom and the stacked bars represent NN1\_Ret and NN10\_Ret respectively. Results suggest that SciBERT embeddings for textual neighbors of scientific text are the most optimal.
    }
    \label{fig:exp5}
\end{figure}

\begin{figure}[!t]
    \centering
    \begin{subfigure}[b]{0.48\columnwidth}
        \centering
        \includegraphics[trim={0.1cm 0.1cm 0 0},clip,width=\textwidth]{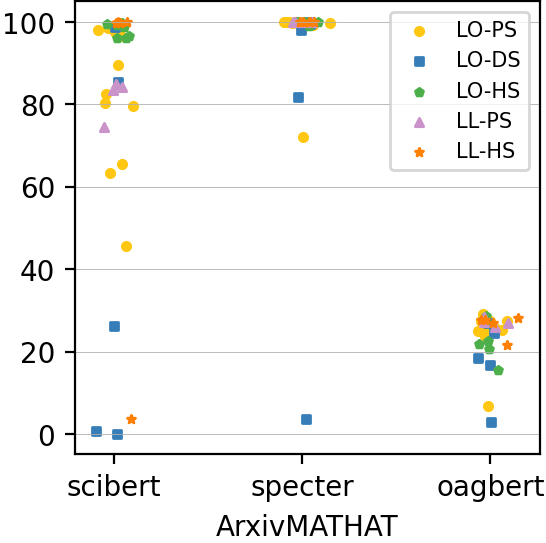}
    \end{subfigure}
    \hfill
    \begin{subfigure}[b]{0.48\columnwidth}
        \centering
        \includegraphics[trim={0.1cm 0.1cm 0 0},clip,width=\textwidth]{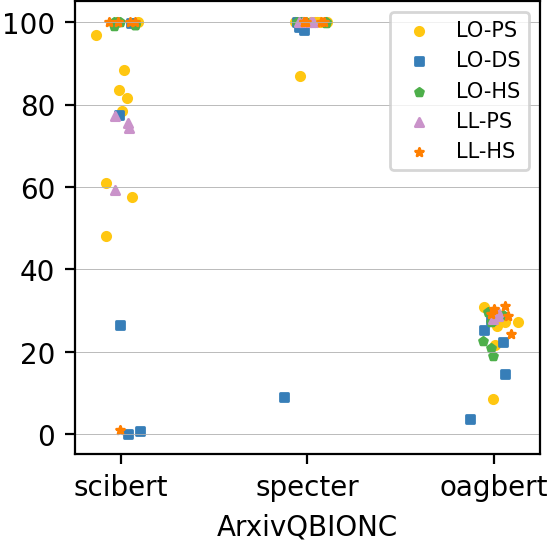}
    \end{subfigure}
    \caption{Distribution of NN1\_Ret for each textual neighbor category. SciBERT embeddings preserve the hierarchy of NN1\_Ret, i.e. PS categories (LO-PS and LL-PS) have lower values than HS categories (LO-HS and LL-HS).
    }
    \label{fig:exp52}
\end{figure}

\subsection{Similarity of Textual Neighbors with Original Documents}
\label{ssec:altret_tneigh}
Let $\mathcal{F}$ = \{$f^1$, $f^2$, ..., $f^n$\} be the set of textual neighbor functions described in Table 1. We query different types of textual neighbor classes against the documents embeddings (T+A). We compute the percentage of queries that successfully rank the original document in the top-1 and top-10 ranked list. We expect \textit{HS} and \textit{PS} categories to rank the original document higher in the rank list, and \textit{DS} to rank it lower. If any of the textual neighbor classes or categories don't show the expected behavior, it can be asserted that the LM is brittle in representing the specific type of textual neighbor. \newline
\textit{\textbf{Query}}:$\mathcal{Q}$ = $\bigcup\limits_{f \in \mathcal{F}}\mathcal{Q}_{f}$ = $\bigcup\limits_{f \in \mathcal{F}}$\{$q_j$: $x_j \in \mathcal{X}$, $q_j$=$f(x_j)$\}\\
\textit{\textbf{Candidate Documents}}: $\mathcal{R}$ = \{$x_k$: $x_k \in \mathcal{X}$\}\\
\textit{\textbf{Task}}: For $q_j \in \mathcal{Q}$, retrieve the most similar documents based on cosine similarity.\\
\textit{\textbf{Evaluation}}: NN1\_Ret and NN1\_10. There is only one relevant document in the candidate set for each $q_j$, which is the corresponding T+A embeddings.\newline 
\begin{table*}[!thb]
\centering
\small{
\resizebox{\hsize}{!}{
\begin{tabular}{|c|l|l|l|l|l|l|l|l|l|l|l|l|l|l|l|}
\hline
\rowcolor{Gray}
\textbf{TN Cat}  & \multicolumn{3}{c|}{\textbf{LO-HS}} & \multicolumn{3}{c|}{\textbf{LO-PS}} & \multicolumn{3}{c|}{\textbf{LO-DS}} & \multicolumn{3}{c|}{\textbf{LL-HS}} & \multicolumn{3}{c|}{\textbf{LL-PS}}      \\ \hline
\rowcolor{Gray}
\textbf{Model} & \multicolumn{1}{c|}{\textbf{SB}} & \multicolumn{1}{c|}{\textbf{SP}} & \multicolumn{1}{c|}{\textbf{OB}} & \multicolumn{1}{c|}{\textbf{SB}} & \multicolumn{1}{c|}{\textbf{SP}} & \multicolumn{1}{c|}{\textbf{OB}} & \multicolumn{1}{c|}{\textbf{SB}} & \multicolumn{1}{c|}{\textbf{SP}} & \multicolumn{1}{c|}{\textbf{OB}} & \multicolumn{1}{c|}{\textbf{SB}} & \multicolumn{1}{c|}{\textbf{SP}} & \multicolumn{1}{c|}{\textbf{OB}} & \multicolumn{1}{c|}{\textbf{SB}} & \multicolumn{1}{c|}{\textbf{SP}} & \multicolumn{1}{c|}{\textbf{OB}} \\ \hline
\textit{ACL-ANTH} & 51.8 & \textbf{68.9} & 4.4 & 20.5 & \textbf{61.8} & 4.3 & 12.0 & 35.6 & \textbf{2.2} & 73.0 & \textbf{81.2} & 5.1 & 13.4 & \textbf{72.0} & 5.2 \\ \hline
\textit{ICLR}  & 52.2 & \textbf{74.6} & 4.7 & 18.6 & \textbf{61.8} & 4.7 & 10.9 & 33.2 & \textbf{2.3} & 71.2 & \textbf{82.2} & 5.5 & 10.8 & \textbf{75.0} & 6.0 \\ \hline
\textit{Arxiv-CS\_SY} & 52.8 & \textbf{74.8} & 5.2 & 23.2 & \textbf{66.0} & 5.1 & 11.6 & 40.9 & \textbf{2.6} & 71.7 & \textbf{83.4} & 6.0 &  14.5 & \textbf{77.6} & 6.6 \\ \hline
\textit{Arxiv-MATH} & 52.2 & \textbf{67.4} & 5.3 & 23.0 & \textbf{62.1} & 5.3  & 14.5 & 34.5 & \textbf{3.2} & 71.1 & \textbf{80.5} & 6.1 & 31.3 & \textbf{75.4} & 6.6  \\ \hline
\textit{Arxiv-ECON}  & 59.0 & \textbf{75.3} & 5.8 & 27.0 & \textbf{66.4} & 5.6  & 14.1 & 40.9 & \textbf{2.8} & 71.2 & \textbf{83.2} & 6.4 & 21.4 & \textbf{76.8} & 6.8 \\ \hline
\textit{Arxiv-QBIO} & 55.2 & \textbf{74.1} & 5.4 & 23.5 & \textbf{65.1} & 4.9  & 12.1 & 38.7 & \textbf{2.6} & 71.0 & \textbf{82.9} & 5.9 & 17.1 & \textbf{76.4} & 6.3 \\ \hline
\textit{Arxiv-HEP} & 53.1 & \textbf{66.2} & 5.9 & 22.9 & \textbf{61.0} & 5.9 & 13.5 & 34.2 & \textbf{3.6} & 71.0 & \textbf{80.0} & 6.8 & 25.4 & \textbf{73.4} & 7.2 \\ \hline
\end{tabular}}
\caption{AOP-10 values for different categories.
The best results for LO-DS category are from OAG-BERT (OB), however that is because the model performs poorly on all categories of textual neighbors. Similarly, best results for the rest four categories are from SPECTER (SP), following which it also has a high overlap percentage for the LO-DS category. SciBERT (SB) embeddings perform the best for HS and DS but falter on PS semantic categories.
}
\label{tab:nn_overlap}
}
\end{table*}
We present the results in Figure~\ref{fig:exp5}. SciBERT and OAG-BERT for the LO-DS category show less than 50\% NN10\_Ret, which is desirable as LO-DS neighbors are semantically dissimilar, and hence should not be neighbors in the embedding space. SciBERT shows improvement via NN10\_Ret over NN1\_Ret for \textit{PS} categories. High NN1\_Ret for \textit{HS} categories indicates SciBERT successfully encodes highly similar texts closer than partially similar texts. OAG-BERT performs poorly for both metrics, indicating that it doesn't encode textual neighbors optimally. SPECTER embeddings perform poorly on the LO-DS category. They however achieve the maximum values showing no difference between LO vs LL, or HS vs PS categories.

To analyse the high values for SPECTER, we present the individual NN1\_Ret for each of the 32 textual neighbor classes in Figure~\ref{fig:exp52} and observe only two classes `TDelNN' and `T\_A\_DelNNChar' which lead to less than 90\% NN1\_Ret. Unlike SPECTER and OAG-BERT, SciBERT preserve the hierarchy, with \textit{Highly Similar} classes ranked higher than \textit{Partially Similar} classes. An interesting case with SciBERT embeddings is that the T\_A\_WS neighbor class belonging to the LL-HS category, has a low NN1\_Ret value across all datasets, suggesting that the SciBERT model is extremely brittle to white space character perturbations (because of the constraint on sequence length). Another breaking point for the SciBERT is the textual neighbor class T\_ARepADJ (replacing adjectives with antonyms) of LO-DS category, which shows high values (around 80\%) for NN1\_Ret which is undesirable. We observe that among the three specific LO-PS categories `T\_ADelQ1', `T\_ADelQ2', and `T\_ADelQ3',  SciBERT performs worst for the `T\_ADelQ3', indicating that the last quantile of the abstract contains relevant information encoded by SciBERT. OAG-BERT shows a reverse trend to SPECTER by achieving low values for all neighbors classes indicating brittleness to text manipulation.

\begin{figure}[!t]
    \centering
    \begin{subfigure}[b]{0.48\columnwidth}
        \centering
        \includegraphics[width=\textwidth]{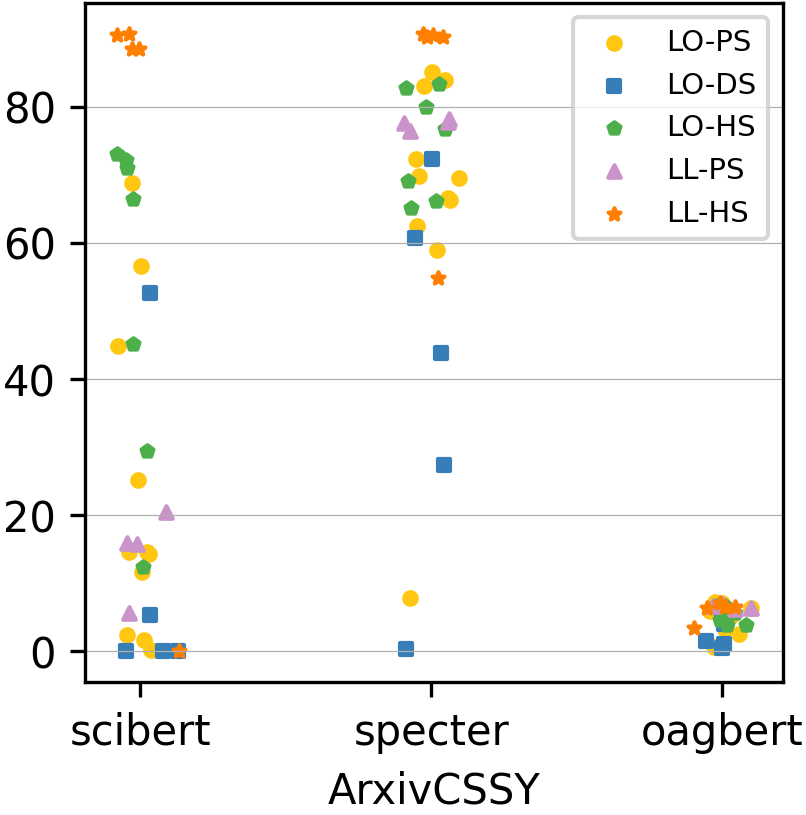}
    \end{subfigure}
    \hfill
    \begin{subfigure}[b]{0.48\columnwidth}
        \centering
        \includegraphics[width=\textwidth]{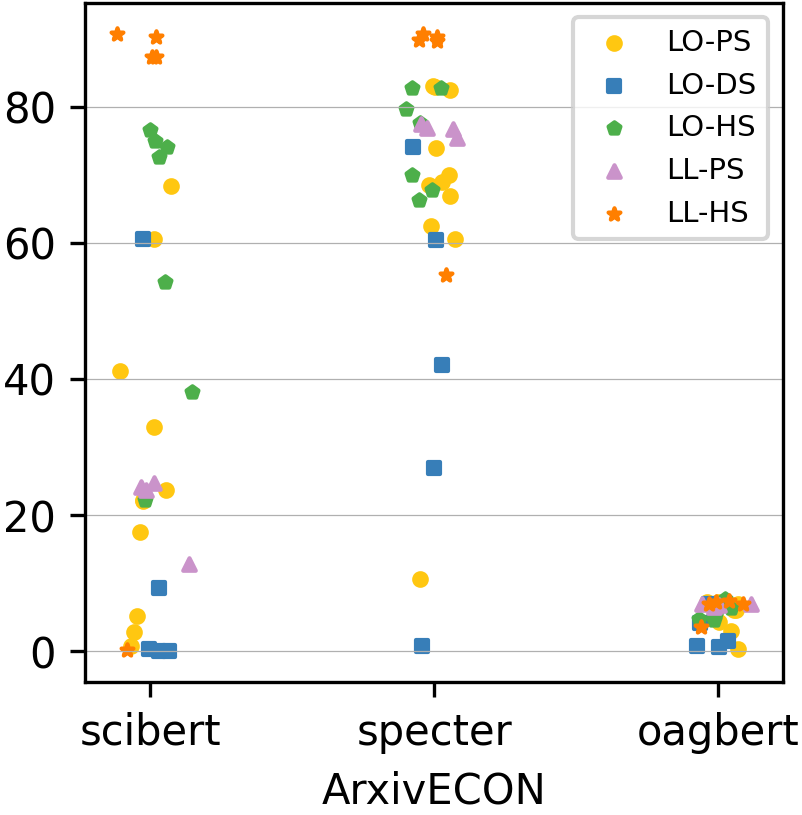}
    \end{subfigure}
    \caption{AOP-10 distribution of all categories of textual neighbors. SciBERT performs poorly for LL-PS (consists of neighbors that scramble abstract sentences). If we ignore LO-DS category, SPECTER embeddings perform decently overall.
    }
    \label{fig:exp6}
\end{figure}

\subsection{Overlap amongst Nearest Neighbors}
We compute the overlap amongst the nearest neighbors of each textual neighbor class and the original document embeddings. 
We randomly sample a query set of 2000 queries for each textual neighbor class (and their corresponding T+A embedding) and compute nearest neighbor (NN) overlap for these. In this task, we are interested in evaluating if NN-based retrieval retrieves the same documents for a textual neighbor class and T+A embedding. \newline
\textit{\textbf{Query}}: $\mathcal{Q}$ = $\mathcal{Q}_f \cup{} \mathcal{Q}_{T+A}$ \\
$\mathcal{Q}_f$ = \{$q_j$: $x_j \in \mathcal{X}_{2000}$, $q_j = f(x_j)$\} \\
$\mathcal{Q}_{T+A}$ = \{$q_j$: $x_j \in \mathcal{X}_{2000}$, $q_j = T+A(x_j)$\} \\
\textit{\textbf{Candidate Documents}}: $\mathcal{R}$ = \{$q_j$: $x_j \in \mathcal{X}$, $q_j$=$f(x_j)$\} $\cup{}$ \{$q_j$: $x_j \in \mathcal{X}$, $f=T+A$, $q_j$=$f(x_j)$\} \\
\textit{\textbf{Task}}: For each pair of $q_j, q_k \in \mathcal{Q}$, such that $q_j \in \mathcal{Q}_f$ and $q_k \in \mathcal{Q}_{T_A}$, compute the overlap among ten nearest neighbors (NN-10) of $q_j$ and $q_k$. \\
\textit{\textbf{Evaluation}}: AOP-10.\\
We present the results arranged by Textual Neighbor categories in Table~\ref{tab:nn_overlap}. The individual AOP-10 distribution of all categories of textual neighbors is presented in Figure~\ref{fig:exp6}. While overall results look good for SPECTER, it should be noted that SPECTER performs poorly for embedding LO-DS textual neighbors indicating that it has a shallow understanding of semantics. AOP-10 values for SPECTER show a positive trend: LL-HS $>$ LO-HS and LL-HS $>$ LL-PS. SciBERT performs decently for the \textit{HS} category, but its performance drops for the \textit{PS} categories. OAG-BERT has the lowest AOP-10 for all textual neighbor categories. When put in perspective against the previous NN1\_Ret and NN10\_Ret, we believe that SciBERT performs decently in encoding the HS and PS neighbors closer to the original T+A embedding. However, low value of AOP-10 for SciBERT for PS neighbors reflects that while the PS neighbors are closer to the original document in comparison to others, the original document has other nearest neighbors than the PS neighbor.

We present a matrix to summarize the capability of the models in Table~\ref{tab:tneigh_matrix} in encoding the five textual neighbor categories. The five categories arranged in increasing order of semantic similarity are: LL-HS $\geq$ LO-HS $>$ LL-PS $>$ LO-PS $>$ LO-DS. We use heuristic-based values to define optimality. For each of the five categories, we define AOP-20 thresholds to classify if the textual neighbor representations for the corresponding are optimal or not. It is expected that AOP-20 values for semantic categories should be in order: HS $>$ PS $>$ DS. AOP-20 values for orthographic categories should follow: LL $>$ LO.
\begin{table}[!t]
\centering
\resizebox{\columnwidth}{!}{
\small{
\begin{tabular}{|l|c|c|c|c|c|}
\hline
\rowcolor{Gray}
 & \textbf{LL-HS} & \textbf{LO-HS} & \textbf{LL-PS} & \textbf{LO-PS} & \textbf{LO-DS} \\ \hline
\textbf{Threshold} & $>$ 75 & $>$ 70 & \multicolumn{2}{l|}{50 $<$ AOP-20 $<$ 70} & $<$ 20 \\ \hline
\textbf{SciBERT} & \cellcolor{Gray}\textbf{75.04} & 59.62 & 22.6 & 26.28 & 14.35 \\ \hline
\textbf{SPECTER} & \cellcolor{Gray}\textbf{86.77} & \cellcolor{Gray}\textbf{77.55} & 80.93 & \cellcolor{Gray}\textbf{69.03} & 41.32  \\ \hline
\textbf{OAG-BERT} & 6.84 & 6.24 & 7.31 & 5.95 & \cellcolor{Gray}\textbf{3.17} \\ \hline
\end{tabular}
}}
\caption{Capability of the models in encoding textual neighbor categories optimally in the embedding space. Gray cells represent optimal representations for each model based on AOP-20.}
\label{tab:tneigh_matrix}
\end{table}

\section{Conclusion}
We propose five categories of textual neighbors to organize the increasing number of textual neighbor types: LL-HS, LO-HS, LL-PS, LO-PS, and LO-DS. We evaluate SciBERT, SPECTER, and OAG-BERT models on thirty-two textual neighbor classes organized into the previous five categories. We show that evaluation of language models on `Semantically Dissimilar' texts is also important rather than just evaluation on `Semantically Similar' texts. We show that the SciBERT model is highly sensitive to the input length. SPECTER embeddings for all types of textual neighbors are highly similar irrespective of whether the textual neighbor is semantically dissimilar or not. SPECTER embeddings show sensitivity to the presence of specific keywords. Lastly, OAG-BERT embeddings of all categories of textual neighbors are highly dissimilar to the original title and abstract (T+A) embeddings. We believe that our insights could be used to develop better pretraining strategies for scientific document language models and also to evaluate other language models. One example for MLM (or replaced token identification) could be utilizing a weighted-penalty based loss, i.e. partially similar tokens if predicted should be penalized less in comparison to the prediction of unrelated (or dissimilar) tokens. Additionally, these insights could also be utilised by several systems that use these scientific document language models to incorporate informed strategies in downstream systems such as recommendation systems.

\bibliography{anthology,custom}
\bibliographystyle{acl_natbib}

\appendix
\section*{Appendix}
\section{Candidate Retrieval}
\label{sec:GSSSCandidates}
The candidates fetched for the queries `document vector' and `document vectors' on Google Scholar and Semantic Scholar in Table~\ref{apptab:gsssresult}. It can be observed that both queries have no common candidates on either of the search engines.
\begin{table*}[]
\newcolumntype{L}{>{\centering\arraybackslash}p{0.98\columnwidth}}
\centering
\small{
\begin{tabular}{|L|L|}
\hline
\rowcolor{Gray}
\textbf{Query = document vector}  & \textbf{Query = document vectors}  \\ \hline
\rowcolor{Gray}
Semantic Scholar & Semantic Scholar \\ \hline
Corporate value evaluation using patent document vector & Using Sparse Composite Document Vectors to Classify VBA Macros \\ \hline
Improving a tf-idf weighted document vector embedding & SCDV : Sparse Composite Document Vectors using soft clustering over distributional representations \\ \hline
Legal Document Retrieval using Document Vector Embeddings and Deep Learning & Music genre classification with word and document vectors \\ \hline
Document vector embeddings for bibliographic records indexing & Constructing Document Vectors Using Kernel Density Estimates \\ \hline
Document Vector Extension for Documents Classification & Text classification with sparse composite document vectors \\ \hline
Multi-document Summarization by Creating Synthetic Document Vector Based on Language Model & Words are not Equal: Graded Weighting Model for Building Composite Document Vectors \\ \hline
Document vector representations for feature extraction in multi-stage document ranking & A Document Descriptor using Covariance of Word Vectors \\ \hline
An Adaptive Topic Tracking Model Based on 3-Dimension Document Vector & Document Embedding with Paragraph Vectors \\ \hline
A support vector machine mixed with TF-IDF algorithm to categorize Bengali document & Document classification with distributions of word vectors \\ \hline
A framework for a feedback process to analyze and personalize a document vector space in a feature extraction model & Text document clustering using global term context vectors \\ \hline
\rowcolor{Gray}
Google Scholar & Google Scholar \\ \hline
Document vector representations for feature extraction in multi-stage document ranking & Document embedding with paragraph vectors \\ \hline
Document ranking and the vector-space model & SCDV: Sparse Composite Document Vectors using soft clustering over distributional representations \\ \hline
Legal document retrieval using document vector embeddings and deep learning & Text document clustering using global term context vectors \\ \hline
Improving a tf-idf weighted document vector embedding & Document classification with distributions of word vectors \\ \hline
Efficient vector representation for documents through corruption & Using sparse composite document vectors to classify vba macros \\ \hline
Enhancing web service clustering using Length Feature Weight Method for service description document vector space representation & Words are not equal: Graded weighting model for building composite document vectors \\ \hline
Document vector compression and its application in document clustering & Image-based document vectors for text retrieval \\ \hline
A vector space model for automatic indexing & Improving Document Vectors Representation using Semantic Links and Attributes \\ \hline
Hierarchical document categorization with support vector machines & Self organization of a massive document collection \\ \hline
Using an n-gram-based document representation with a vector processing retrieval model & Music genre classification with word and document vectors \\ \hline
\end{tabular}
}
\caption{Candidate documents retrieved for queries `document vector' and `document vectors'.}
\label{apptab:gsssresult}
\end{table*}

\section{Textual Neighbors}
\label{sec:app_tneigh}
We present examples for textual neighbors in Table~\ref{apptab:examples_tneig}. We use NLTK for preprocessing text and constructing textual neighbors.

\begin{table*}
\newcolumntype{K}{>{\arraybackslash}p{0.4\columnwidth}}
\newcolumntype{L}{>{\arraybackslash}p{1.3\columnwidth}}
\centering
\small{
\resizebox{\hsize}{!}{
\begin{tabular}{p{0.1\columnwidth}p{0.25\columnwidth}p{0.4\columnwidth}p{1.3\columnwidth}}
\rowcolor{Gray}
\multicolumn{1}{l}{\textbf{O-S}} &
\multicolumn{1}{l}{  \begin{tabular}[c]{@{}l@{}}\textbf{Neighbor} \\ \textbf{Code}\end{tabular}  } & \multicolumn{1}{l}{\textbf{Form}} & \multicolumn{1}{l}{\textbf{Example}} \\ \hline
LL-PS & T\_ARot & \begin{tabular}[K]{@{}K@{}}T $\rightarrow$ Preserve\\ A $\rightarrow$ Rotate\end{tabular} & \begin{tabular}[L]{@{}L@{}}\textbf{A: }We introduce a representation for computer programs based on language models. We train deep robust embeddings using pytorch. Contextual embeddings are common in NLP.\end{tabular} \\ \hline
LL-PS & T\_AShuff & \begin{tabular}[c]{@{}l@{}}T $\rightarrow$ Preserve\\ A $\rightarrow$ Shuffle\end{tabular} &  \begin{tabular}[L]{@{}L@{}}\textbf{A: }Contextual embeddings are common in NLP. We introduce a representation for computer programs based on language models. We train deep robust embeddings using pytorch.\end{tabular} \\ \hline
LL-PS & T\_ASortAsc & \begin{tabular}[p]{@{}l@{}}T $\rightarrow$ Preserve\\ A $\rightarrow$ Sort Ascending\end{tabular} & \begin{tabular}[L]{@{}L@{}} \textbf{A: }Contextual embeddings are common in NLP. We train deep robust embeddings using pytorch. We introduce a representation for computer programs based on language models.\end{tabular} \\ \hline
LL-PS & T\_ASortDesc & \begin{tabular}[c]{@{}l@{}}T $\rightarrow$ Preserve\\ A $\rightarrow$ Sort Descending\end{tabular} & \begin{tabular}[L]{@{}L@{}}\textbf{A: }We introduce a representation for computer programs based on language models. We train deep robust embeddings using pytorch. Contextual embeddings are common in NLP.\end{tabular} \\ \hline
LO-PS & T\_ADelRand & \begin{tabular}[c]{@{}l@{}}T $\rightarrow$ Preserve\\ A $\rightarrow$ Random word \\ deletion 30\%\end{tabular} & \begin{tabular}[L]{@{}L@{}} \textbf{A: }Contextual  are common in .  introduce a representation  computer programs based on language models. We train deep robust  using . \end{tabular} \\ \hline
LO-PS & T\_ADelADJ & \begin{tabular}[c]{@{}l@{}}T $\rightarrow$ Preserve\\ A $\rightarrow$ Delete all ADJs\end{tabular} & \begin{tabular}[L]{@{}L@{}} \textbf{A: } We introduce a representation for computer programs based on language models . We train embeddings using pytorch .\end{tabular} \\ \hline
LO-DS & T\_ADelNN & \begin{tabular}[c]{@{}l@{}}T $\rightarrow$ Preserve\\ A $\rightarrow$ Delete all NNs\end{tabular} & \begin{tabular}[L]{@{}L@{}}\textbf{A: }Contextual are common in . We introduce a for based on . We train deep robust using .\end{tabular} \\ \hline
LO-PS & T\_ADelVB & \begin{tabular}[c]{@{}l@{}}T $\rightarrow$ Preserve\\ A $\rightarrow$ Delete all Verbs\end{tabular} & \begin{tabular}[L]{@{}L@{}}\textbf{A: }Contextual embeddings common in NLP . We a representation for computer programs on language models . We deep robust embeddings pytorch .\end{tabular} \\ \hline
LO-PS & T\_ADelADV & \begin{tabular}[c]{@{}l@{}}T $\rightarrow$ Preserve\\ A $\rightarrow$ Delete all ADVs\end{tabular} & \begin{tabular}[L]{@{}L@{}}\textbf{A: }Contextual embeddings are common in NLP . We introduce a representation for computer programs based on language models . We train deep robust embeddings using pytorch .\end{tabular} \\ \hline
LO-PS & T\_ADelPR & \begin{tabular}[c]{@{}l@{}}T $\rightarrow$ Preserve\\ A $\rightarrow$ Delete all PRs\end{tabular} & \begin{tabular}[L]{@{}L@{}} \textbf{A: }Contextual embeddings are common in NLP . introduce a representation for computer programs based on language models . train deep robust embeddings using pytorch .\end{tabular} \\ \hline
LO-HS & T\_ADelDT & \begin{tabular}[c]{@{}l@{}}T $\rightarrow$ Preserve\\ A $\rightarrow$ Delete all DTs\end{tabular} & \begin{tabular}[L]{@{}L@{}} \textbf{A: }Contextual embeddings are common in NLP . We introduce representation for computer programs based on language models . We train deep robust embeddings using pytorch .\end{tabular} \\ \hline
LO-PS & T\_ADelNum & \begin{tabular}[c]{@{}l@{}}T $\rightarrow$ Preserve\\ A $\rightarrow$ Delete all Numbers\end{tabular} & \begin{tabular}[L]{@{}L@{}} \textbf{A: }Contextual embeddings are common in NLP . We introduce a representation for computer programs based on language models . We train deep robust embeddings using pytorch .\end{tabular} \\ \hline
LO-DS & T\_ADelNNPH & \begin{tabular}[c]{@{}l@{}}T $\rightarrow$ Preserve\\ A $\rightarrow$ Delete  all NN \\ Phrases\end{tabular} & \begin{tabular}[L]{@{}L@{}} \textbf{A: }Contextual embeddings are common in NLP. We introduce a representation for  based on . We train deep  using pytorch.\end{tabular} \\ \hline
LO-PS & T\_ADelTopNNPH & \begin{tabular}[c]{@{}l@{}}T $\rightarrow$ Preserve\\ A $\rightarrow$ Delete top 50\% NPs\end{tabular} & \begin{tabular}[L]{@{}L@{}} \textbf{A: }Contextual embeddings are common in NLP. We introduce a representation for  based on . We train deep robust embeddings using pytorch.\end{tabular} \\ \hline
LO-HS & TDelADJ\_A & \begin{tabular}[c]{@{}l@{}}T $\rightarrow$ Delete all ADJs\\ A $\rightarrow$ Preserve\end{tabular} & \begin{tabular}[L]{@{}L@{}} \textbf{T: }Source Code Embeddings from Language Models \end{tabular} \\ \hline
LO-HS & TDelNN\_A & \begin{tabular}[c]{@{}l@{}}T $\rightarrow$ Delete all NNs\\ A $\rightarrow$ Preserve\end{tabular} & \textbf{T: }from \\ \hline
LO-HS & TDelVB\_A & \begin{tabular}[c]{@{}l@{}}T $\rightarrow$ Delete all VBs\\ A $\rightarrow$ Preserve\end{tabular} & \textbf{T: }Source Code Embeddings from Language Models \\ \hline
LO-HS & TDelDT\_A & \begin{tabular}[c]{@{}l@{}}T $\rightarrow$ Delete all DTs\\ A $\rightarrow$ Preserve\end{tabular} & \textbf{T: }Source Code Embeddings from Language Models \\ \hline
LO-DS & TDelNN & \begin{tabular}[c]{@{}l@{}}T $\rightarrow$ Delete all NNs\\ A $\rightarrow$ Delete\end{tabular} & \textbf{T: }from\\ \hline
LO-PS & T\_ADelQ1 & \begin{tabular}[c]{@{}l@{}}T $\rightarrow$ Preserve\\ A $\rightarrow$ Delete quantile 1\end{tabular} & \textbf{A: }We introduce a representation for computer programs based on language models. We train deep robust embeddings using pytorch. \\ \hline
LO-PS & T\_ADelQ2 & \begin{tabular}[c]{@{}l@{}}T $\rightarrow$ Preserve\\ A $\rightarrow$ Delete quantile 2\end{tabular} &  \textbf{A: }Contextual embeddings are common in NLP. We train deep robust embeddings using pytorch. \\ \hline
LO-PS & T\_ADelQ3 & \begin{tabular}[c]{@{}l@{}}T $\rightarrow$ Preserve\\ A $\rightarrow$ Delete quantile 3\end{tabular} &  \begin{tabular}[L]{@{}L@{}} \textbf{A: }Contextual embeddings are common in NLP. We introduce a representation for computer programs based on language models.\end{tabular} \\ \hline
LL-HS & TNNU\_A & \begin{tabular}[K]{@{}K@{}}T $\rightarrow$ Uppercase NNs\\ A $\rightarrow$ Preserve\end{tabular} & \textbf{T: }SOURCE CODE EMBEDDINGS from LANGUAGE MODEL\\ \hline
LL-HS & TNonNNU\_A & \begin{tabular}[c]{@{}l@{}}T $\rightarrow$ Uppercase non NNs\\ A $\rightarrow$ Preserve\end{tabular} & \textbf{T: }Source Code Embeddings FROM Language Models\\ \hline
LL-HS & T\_ANNU & \begin{tabular}[c]{@{}l@{}}T $\rightarrow$ Preserve\\ A $\rightarrow$ Uppercase NNs\end{tabular} & \begin{tabular}[L]{@{}L@{}}\textbf{A: }Contextual EMBEDDINGS are common in NLP . We introduce a REPRESENTATION for COMPUTER PROGRAMS based on LANGUAGE MODELS . We train deep robust EMBEDDINGS using PYTORCH .\end{tabular} \\ \hline
\end{tabular}}}
\end{table*}

\begin{table*}
\centering
\newcolumntype{K}{>{\arraybackslash}p{0.4\columnwidth}}
\newcolumntype{L}{>{\arraybackslash}p{1.3\columnwidth}}
\small{
\resizebox{\hsize}{!}{
\begin{tabular}{p{0.1\columnwidth}p{0.25\columnwidth}p{0.4\columnwidth}p{1.3\columnwidth}}
\rowcolor{Gray}
\multicolumn{1}{l}{\textbf{O-S}} &
\multicolumn{1}{l}{  \begin{tabular}[c]{@{}l@{}}\textbf{Neighbor} \\ \textbf{Code}\end{tabular}  } & \multicolumn{1}{l}{\textbf{Form}} & \multicolumn{1}{l}{\textbf{Example}} \\ \hline
LL-HS & T\_ANonNNU & \begin{tabular}[K]{@{}K@{}}T $\rightarrow$ Preserve\\ A $\rightarrow$ Uppercase non NNs\end{tabular} & \begin{tabular}[L]{@{}L@{}} \textbf{A: }CONTEXTUAL embeddings ARE COMMON IN NLP . WE INTRODUCE A representation FOR computer programs BASED ON language models . WE TRAIN DEEP ROBUST embeddings USING pytorch .  \end{tabular} \\ \hline
LO-PS & T\_A\_DelNNChar & \begin{tabular}[K]{@{}K@{}}T $\rightarrow$ Delete chars from NNs\\ A $\rightarrow$ Delete chars from NNs\end{tabular} & \begin{tabular}[L]{@{}L@{}} \textbf{T: }Soue Cod Emddings from Lguage dels \\ \textbf{A: }Contextual mbedding are common in NLP . We introduce a representaio for cmuter prrams based on lngage modl . We train deep robust emeddngs using ytorh . \end{tabular} \\ \hline
LO-HS & TRepNNT\_A & \begin{tabular}[K]{@{}K@{}}T $\rightarrow$ Add a NN from title\\ A $\rightarrow$ Preserve\end{tabular} & \begin{tabular}[L]{@{}L@{}} \textbf{T: }Source Source Code Embeddings from Language Models \end{tabular} \\ \hline
LO-HS & TRepNNA\_A & \begin{tabular}[K]{@{}K@{}}T $\rightarrow$ Add a NN from abs\\ A $\rightarrow$ Preserve\end{tabular} & \begin{tabular}[L]{@{}L@{}} \textbf{T: }Source Code Embeddings from Language Models embeddings NLP representation computer \end{tabular} \\ \hline
LO-DS & T\_ADelNonNNs & \begin{tabular}[K]{@{}K@{}}T $\rightarrow$ Preserve\\ A $\rightarrow$ Delete all non NNs\end{tabular} & \begin{tabular}[L]{@{}L@{}} \textbf{A: }NLP representation computer programs language models embeddings pytorch \end{tabular} \\ \hline
LO-DS & T\_ARepADJ & \begin{tabular}[K]{@{}K@{}}T $\rightarrow$ Preserve\\ A $\rightarrow$ Replace ADJs \\ with antonyms\end{tabular} & \begin{tabular}[L]{@{}L@{}} \textbf{A: }Contextual embeddings are individual in NLP . We introduce a representation for computer programs based on language models . We train shallow frail embeddings using pytorch . \end{tabular} \\ \hline
LL-HS & T\_A\_WS & \begin{tabular}[K]{@{}K@{}}Randomly replace 50\% whitespace chars with 2-5 whitespace chars in T \& A\end{tabular} & \begin{tabular}[L]{@{}L@{}} \textbf{TA: }
Source\_Code\_\_\_Embeddings\_from\_\_\_\_Language\_Models.\_Contextual\_ embeddings\_are\_common\_in\_NLP.\_We\_introduce\_a\_representation\_for \_computer\_programs\_based\_on\_language\_models.\_We\_train\_deep\_robust\_ embeddings\_\_\_using\_\_\_\_pytorch. [WS represented using \_]
\end{tabular} \\ \hline
\end{tabular}}}
\caption{Neighbor code is in format Txx\_Ayy\_zz, where xx and yy denote the perturbation to the paper title (T) and the abstract (A) respectively. zz denotes perturbation to both T and A. Missing T or A denotes that the corresponding input field is deleted completely.}
\label{apptab:examples_tneig}
\end{table*}

\section{Summary of Scientific LMs}
\label{sec:app_scilm}
We discuss some popular scientific document language models which leverage the transformer architecture.

\textbf{SciBERT}~\citep{beltagy-etal-2019-scibert} is also a BERT model trained on large amounts of scientific data. It is trained on a random sample
of 1.14M papers from the Semantic Scholar Corpus. The training corpus consists of 18\% papers from the computer science domain and 82\% from the broad biomedical domain. Full texts of the papers are used for training. 

\textbf{SPECTER}~\citep{specter2020cohan} uses citation-informed Transformers to generate general-purpose vector representations of scientific documents. Unlike traditional models, they also leverage inter-document relatedness to learn general purpose embeddings that are effective across a variety of downstream tasks without task-specific fine-tuning. SPECTER leverages citations as a signal for document-relatedness and formulate this into a triplet-loss pre-training objective. SPECTER achieves state-of-the-art results on six out of seven document-level tasks for scientific literature in the SCIDOCS~\citep{specter2020cohan} benchmark suite.

\textbf{OAG-BERT}~\citep{liu2021oag} jointly model texts (title and abstract of the paper) and heterogeneous academic entities (authors, research field, venues, and affiliations) to learn representations for a scientific document. The architecture is similar to BERT, however the authors employ multiple techniques to learn entity embeddings. To distinguish different textual and academic entities use entity type embeddings to indicate the entity type. They design an entity aware 2D-positional encoding to indicate the inter-entity and the intra-entity sequence order. 
It also proposes span-aware entity masking to preserve the sequential relationship between the entity's tokens.

\textbf{BioBERT}~\citep{lee2020biobert} is a BERT model pre-trained on large-scale biomedical corpora. BioBERT model is initialized with BERT~cite weights and then pre-trained on PubMed abstracts and PMC full-text articles. 

\noindent Succeeding the BioBERT model, several models have been trained exclusively for Biomedical texts such as ClinicalBERT~\citep{DBLP:journals/corr/abs-1904-05342}, MIMIC-BERT~\citep{si2019enhancing}, PubMedBERT~\citep{gu2020domain}, and BioMegatron~\citep{shin-etal-2020-biomegatron} to list a few. However, in this work, we focus on general purpose scientific language models that have been trained on scientific documents from diverse research fields.

We summarize the details of the SciBERT, SPECTER, and the OAG-BERT model in Table~\ref{apptab:xbertcomp}.
\newcolumntype{a}{>{\columncolor{Gray}}p{0.2\linewidth}}
\begin{table*}[!htb]
\centering
\small{
\resizebox{\hsize}{!}{
\begin{tabular}{ap{0.25\linewidth}p{0.25\linewidth}p{0.25\linewidth}}
\hline
\rowcolor{Gray}
\textbf{}  & \textbf{OAG-BERT}  & \textbf{SPECTER}  & \textbf{SciBERT}  \\ \hline
\textbf{Model Architecture}  & Entity augmented BERT-base &  BERT-base  & BERT-base \\ \hline
\textbf{Model Initialization}  & -  & SciBERT  & -  \\ \hline
\textbf{Loss Function}  & Hybrid Cross Entropy  & Triplet Margin Loss & Cross Entropy  \\ \hline
\textbf{Training corpus}  & Open Academic Graph & Semantic Scholar Corpus  & Semantic Scholar Corpus  \\ \hline
\textbf{Vocabulary}  & OAG-BERT Vocab  & SciVocab  & SciVocab  \\ \hline
\textbf{Vocabulary Size}  & 44,000 tokens  & 30,000 tokens  & 30,000 tokens  \\ \hline
\textbf{Tokenizer} & WordPiece  & WordPiece  & WordPiece  \\ \hline
\textbf{Text Features} & Title, Abstract, Body, FoS, Authors, Venues, Affiliations & Title, Abstract  & Full-text  \\ \hline
\end{tabular}
}}
\caption{Comparison of different transformer based language models for scientific literature}
\label{apptab:xbertcomp}
\end{table*}

\subsection{Non Transformer-based Models}
Majority of Non Transformer based models utilise the Paragraph Vector~\citep{le2014distributed} technique to learn vectors for the textual content. Citation networks are utilised to learn similar embeddings for related papers. Paper2Vec~\citep{ganguly2017paper2vec} learns embeddings by applying DeepWalk~\citep{Perozzi2014DeepWalkOL} on an augmented citation network of papers. Apart from connecting cited papers, the augmented network also connects k nearest neighbors (from textual embeddings generated using Paragraph Vector~\citep{le2014distributed}). Paper2Vec~\citep{DBLP:journals/corr/TianZ17} learns distributed vertex embeddings from matrix factorization on the weighted context definition of nodes. Following a similar technique as Paper2Vec~\citep{ganguly2017paper2vec}, VOPRec (Vector Representation Learning of Papers with Text Information and Structural Identity for Recommendation)~\citep{Kong2021VOPRecVR} learns embeddings from the text using Paragraph Vector~\citep{le2014distributed} and the citation network using Struc2Vec~\cite{DBLP:journals/corr/FigueiredoRS17}. 

\citet{Zhu2019RepresentingAF} present a method to learn scholar paper embeddings (Represent Anything from Scholar Papers) from different scholar entities such as title, authors, publication venue, and citations. It trains the model by trying to maximize the likelihood of references of a paper. It uses an encoder-decoder framework to learn representations from title words, author names, publication venue, and publication year. The proposed method can generate representations for papers even if the references are missing as that information is already encoded in the entities during the training.

As the pretrained models or code for none of the Non Transformer-based models is publicly available, we skip their evaluation in this work.

\section{Analysing Embeddings for Scientific Document Titles and Abstracts}
\label{sec:app_embscidoc}
\begin{figure*}[!thb]
    \setlength{\tabcolsep}{3pt}
    \centering
    \begin{tabular}{ccc} 
        \underline{SciBERT} & \underline{SPECTER} & \underline{OAG-BERT} \\
        Arxiv-QBIO & Arxiv-QBIO & Arxiv-QBIO \\
        \includegraphics[width=0.32\linewidth]{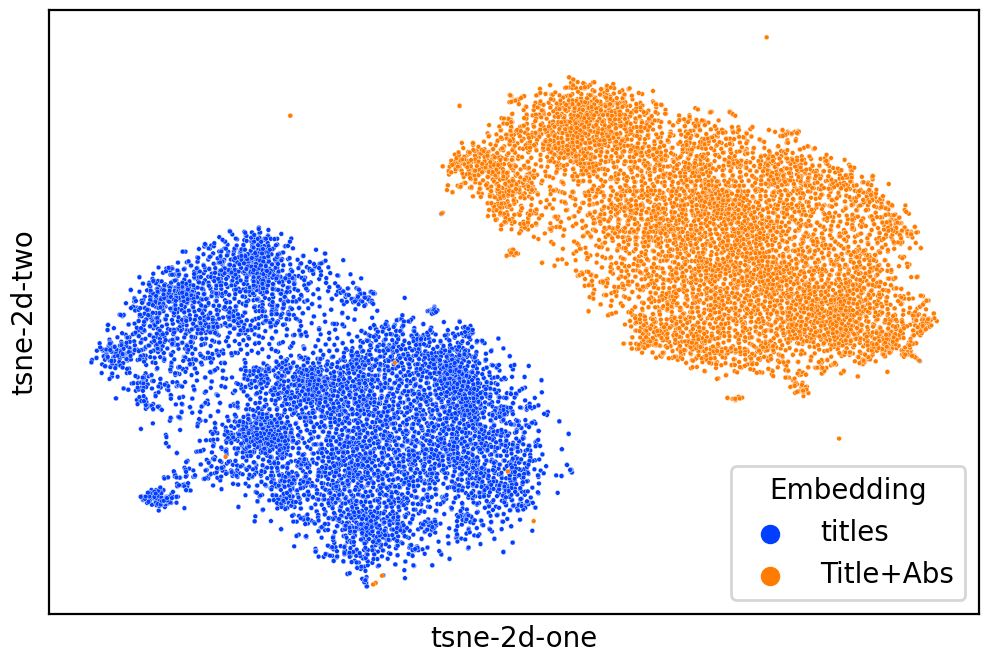} & \includegraphics[width=0.32\linewidth]{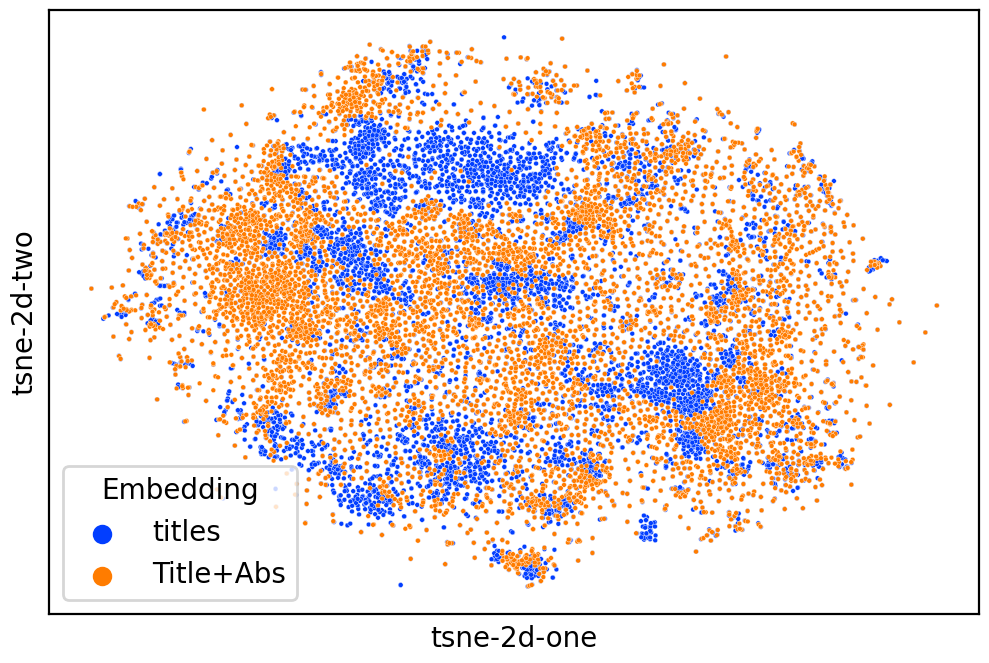} & \includegraphics[width=0.32\linewidth]{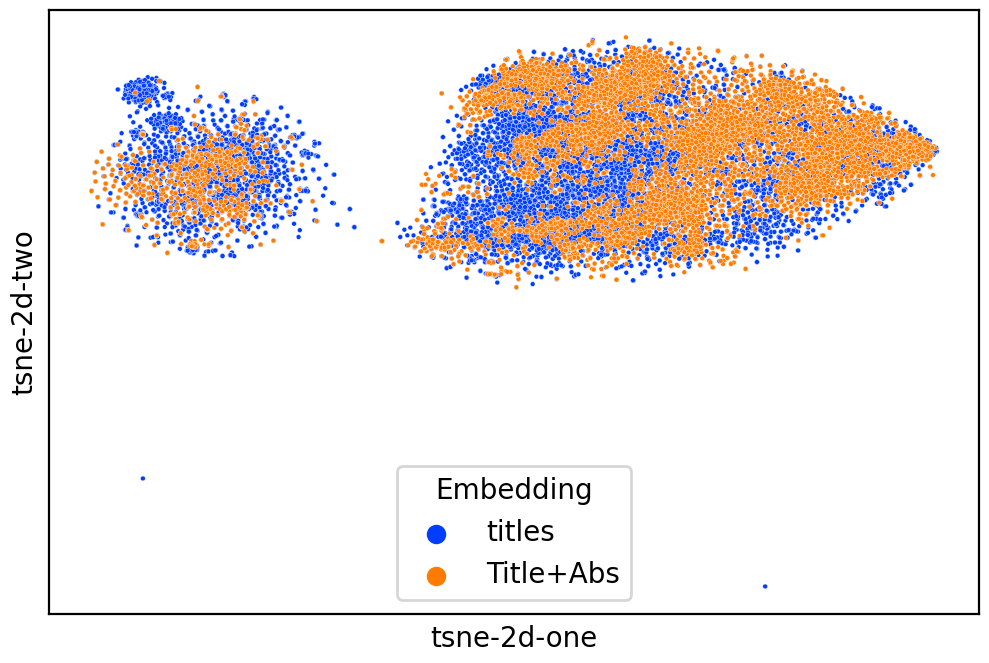} \\
        Arxiv-ECON & Arxiv-ECON & Arxiv-ECON \\
        \includegraphics[width=0.32\linewidth]{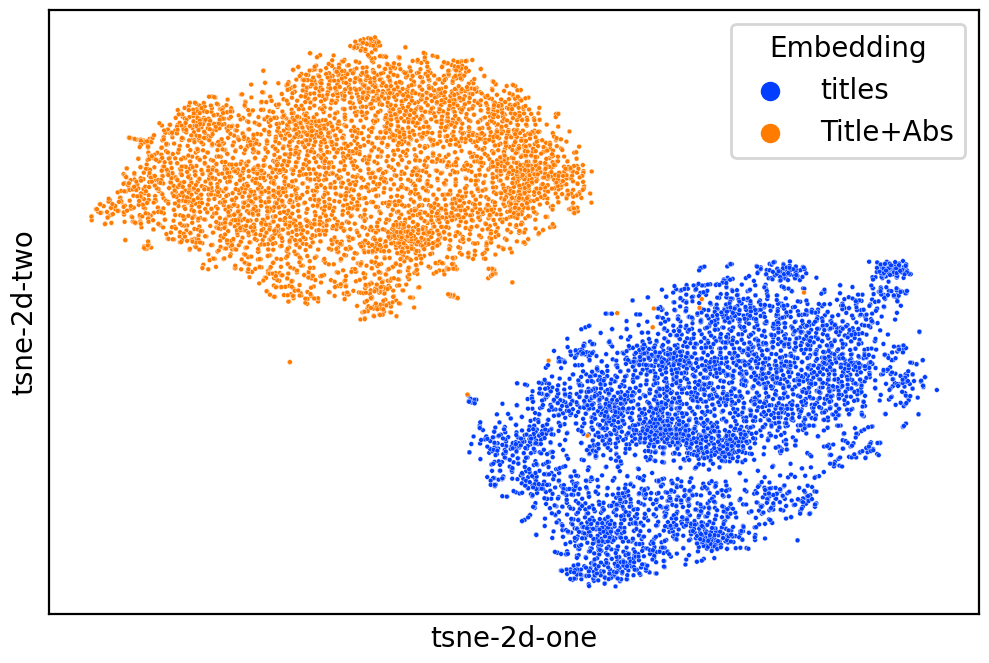} & \includegraphics[width=0.32\linewidth]{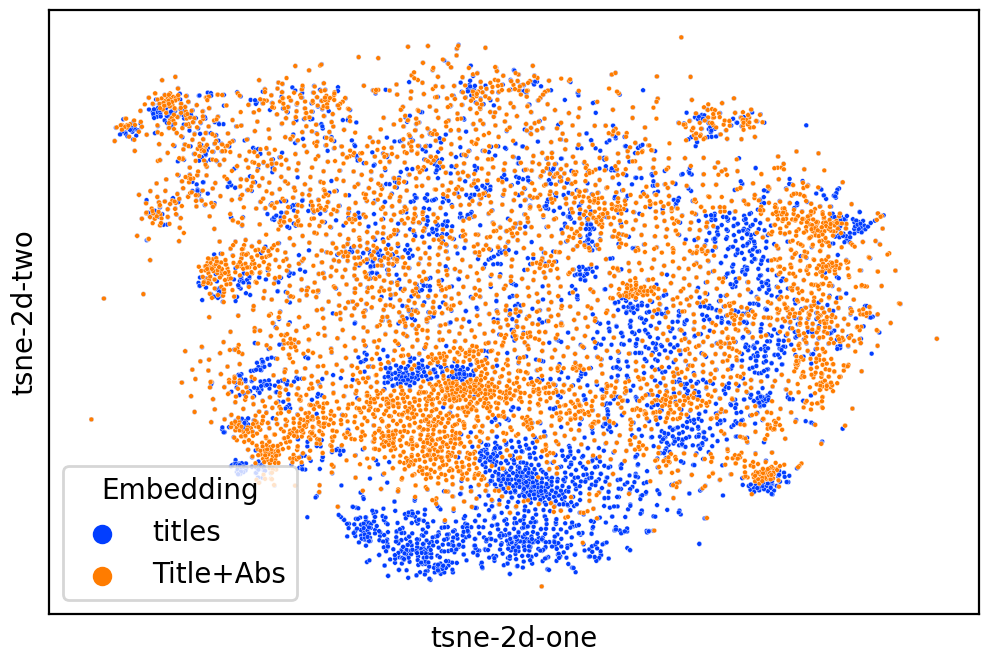} & \includegraphics[width=0.32\linewidth]{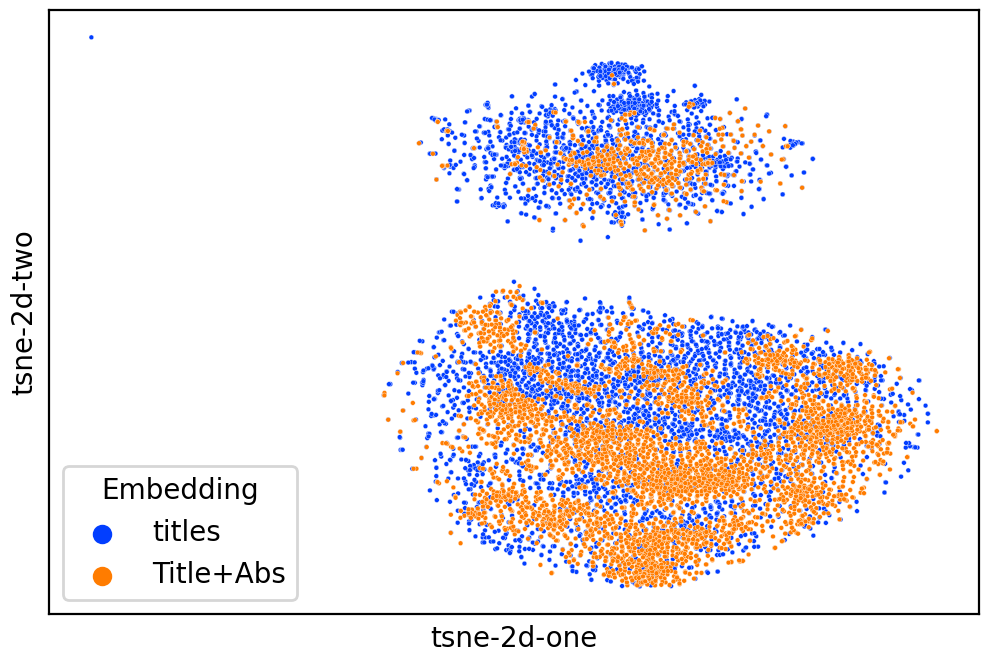} \\
        Arxiv-MATH & Arxiv-MATH & Arxiv-MATH \\
        \includegraphics[width=0.32\linewidth]{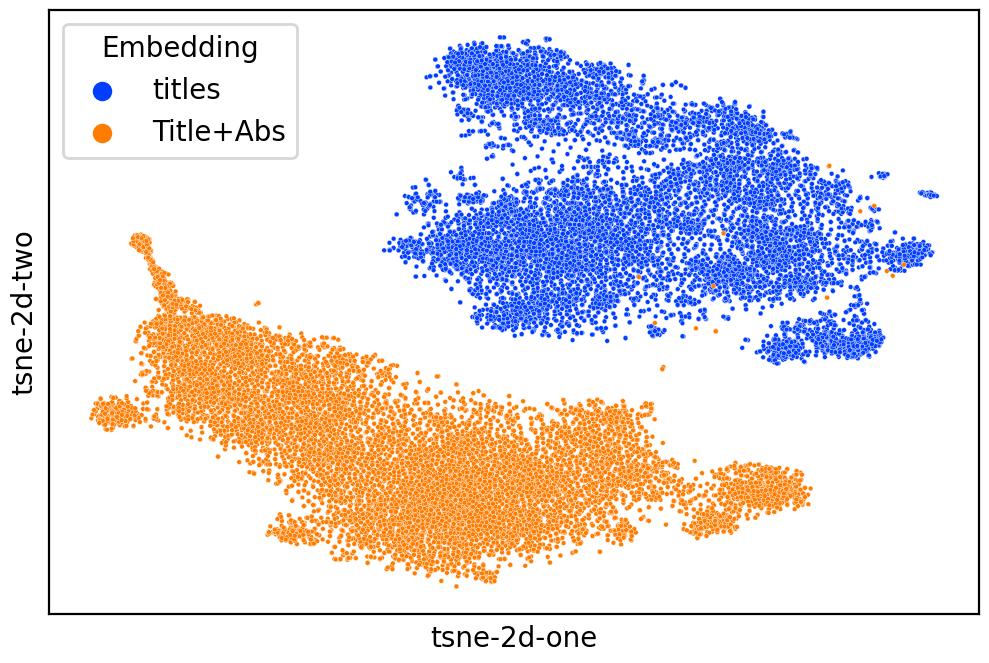} & \includegraphics[width=0.32\linewidth]{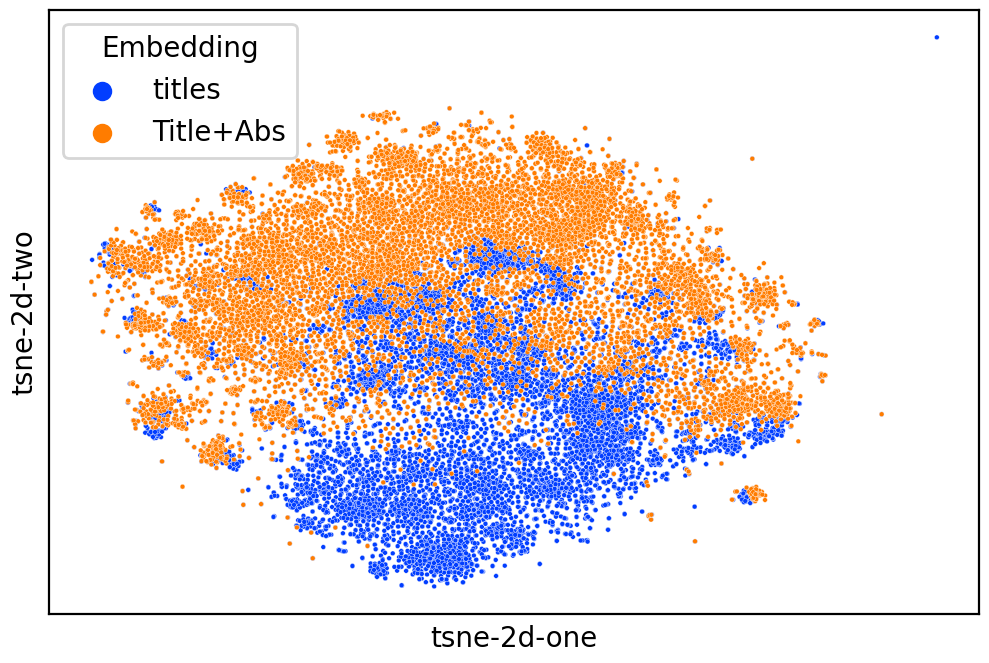} & \includegraphics[width=0.32\linewidth]{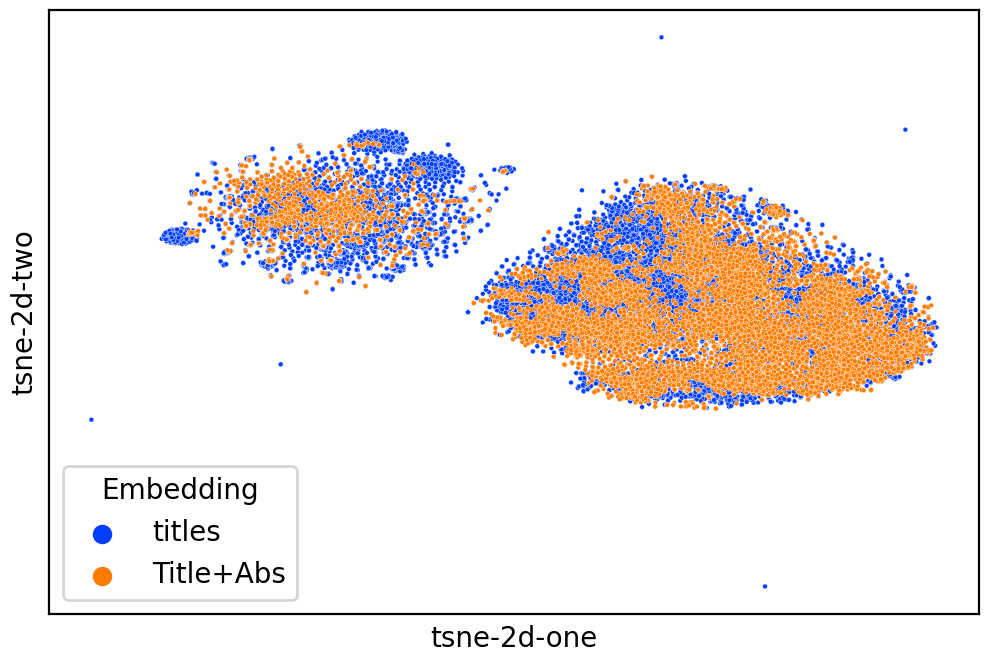} \\
        Arxiv-HEP & Arxiv-HEP & Arxiv-HEP \\
        \includegraphics[width=0.32\linewidth]{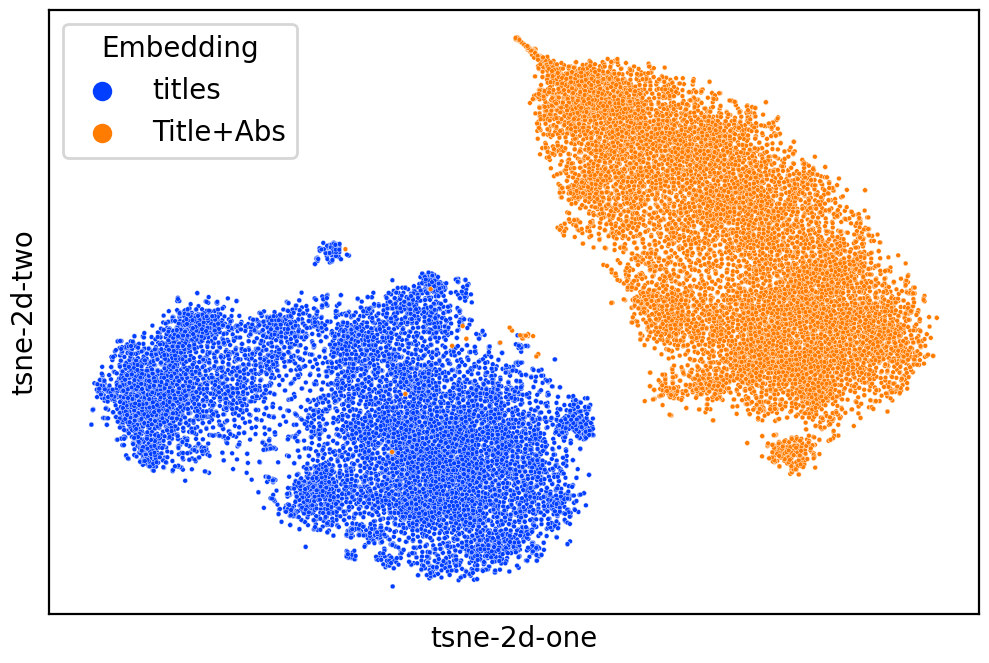} & \includegraphics[width=0.32\linewidth]{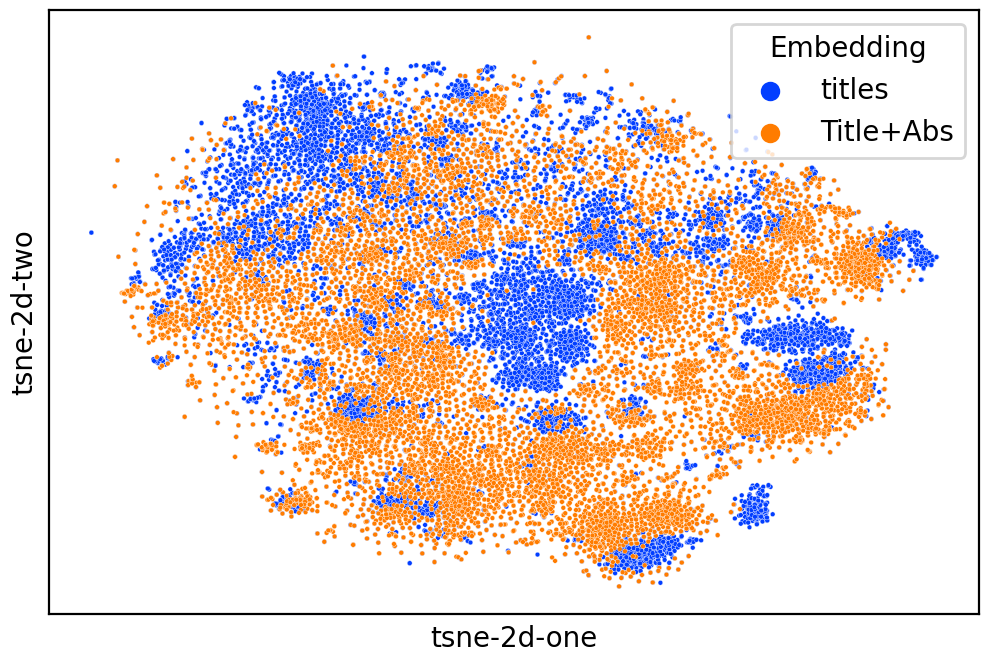} & \includegraphics[width=0.32\linewidth]{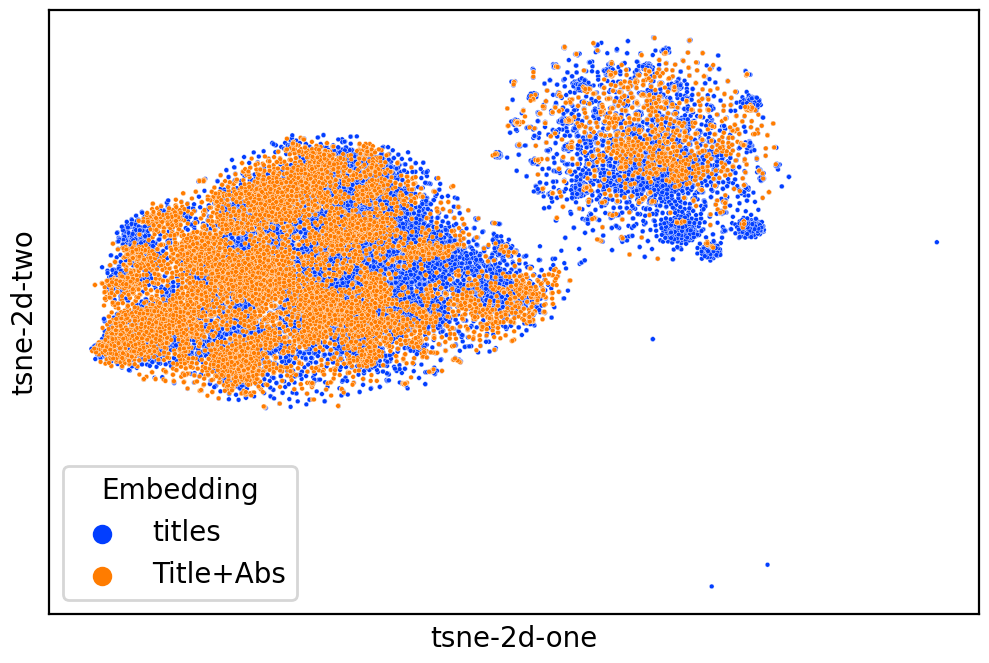} \\
        ACL-ANTH & ACL-ANTH & ACL-ANTH \\
        \includegraphics[width=0.32\linewidth]{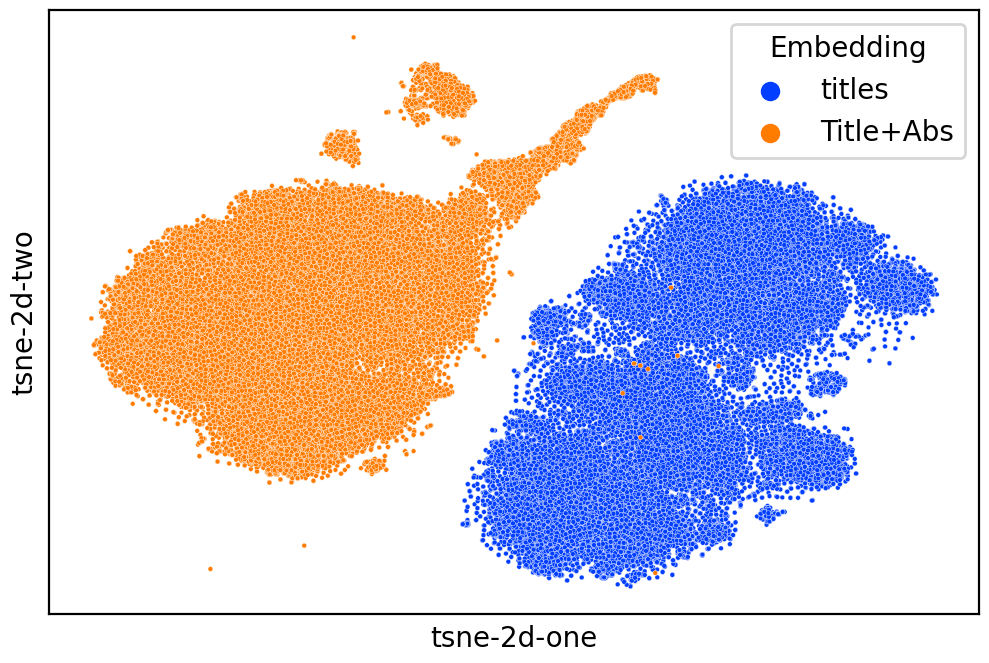} & \includegraphics[width=0.32\linewidth]{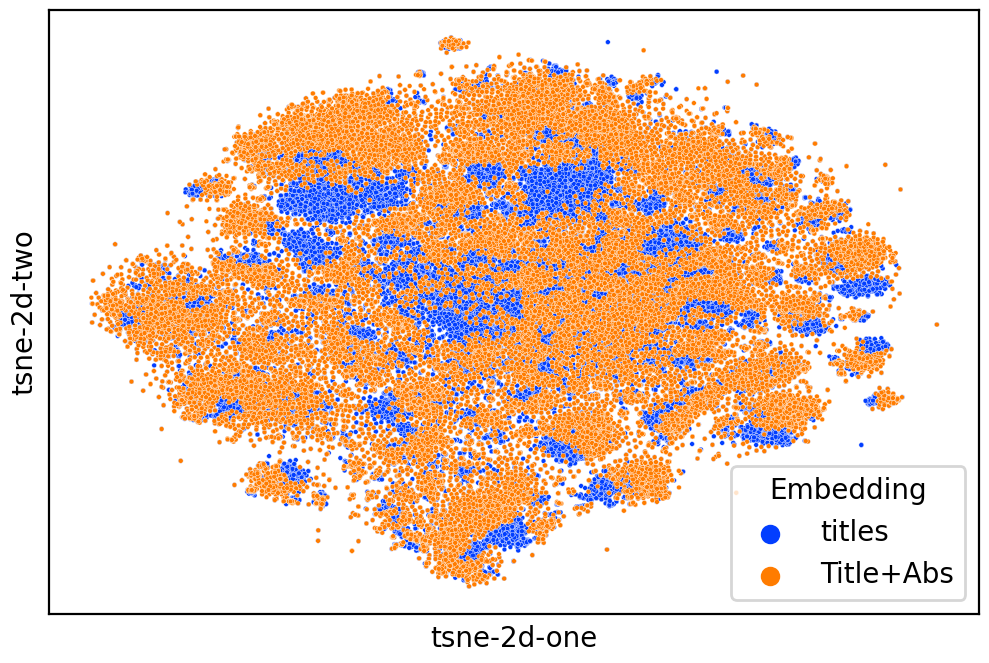} & \includegraphics[width=0.32\textwidth]{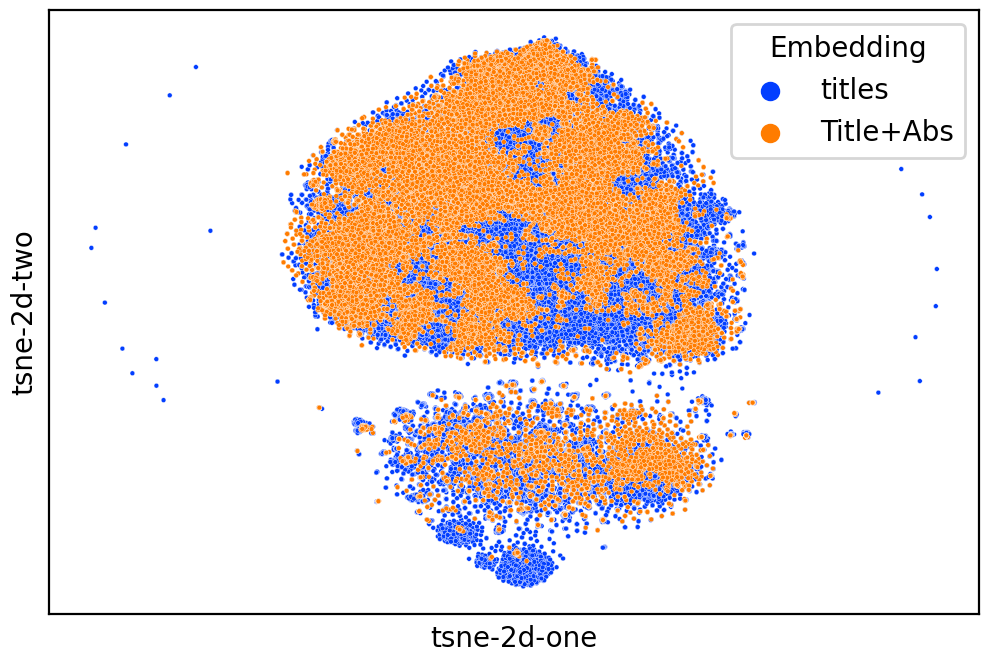} \\
    \end{tabular}
    \caption{t-SNE plots for T and T+A embeddings for various dataset. Completely non-overlapping embeddings for T and T+A by SciBERT model highlight differences in encoding texts of varying lengths.}
    \label{appfig:e2_tsne}
\end{figure*}

We present the t-SNE plots for the Task II in Figure~\ref{appfig:e2_tsne}. The embedding space contains vectors for titles and the T+A embeddings.

\subsection{Normalized Embeddings}
\label{ssec:app_norm_emb_results}
\citet{timkey2021bark} indicate that cosine similarity is dominated by few rouge dimensions in embeddings. We repeat Task I with: (i) L2-normalized embeddings, and (ii) standardization procedure by~\citet{timkey2021bark}. While L2-normalization leads to an incremental improvement, standardization leads to improvement only for SciBERT and SPECTER models. We present the results of \textbf{Task I} (Section~\ref{sec:short-text}) with normalized embeddings in Table~\ref{apptab:normexp1exp2}.
\begin{table*}[!tb]
\centering
\small{
\begin{tabular}{l||cc|cc|cc||cc|cc|cc}
\hline
& \multicolumn{6}{c||}{\textbf{L2-normalized embeddings}} & \multicolumn{6}{c}{\textbf{ Standardization}~\citep{timkey2021bark}} \\ \hline
& \multicolumn{2}{c|}{\textbf{SciBERT}} & \multicolumn{2}{c|}{\textbf{SPECTER}} & \multicolumn{2}{c||}{\textbf{OAG-BERT}} & \multicolumn{2}{c|}{\textbf{SciBERT}} & \multicolumn{2}{c|}{\textbf{SPECTER}} & \multicolumn{2}{c}{\textbf{OAG-BERT}} \\ \hline
& \textit{\textbf{MRR}} & \textit{\textbf{T100}} & \textit{\textbf{MRR}} & \textit{\textbf{T100}} & \textit{\textbf{MRR}} & \textit{\textbf{T100}} & \textit{\textbf{MRR}} & \textit{\textbf{T100}} & \textit{\textbf{MRR}} & \textit{\textbf{T100}} & \textit{\textbf{MRR}} & \textit{\textbf{T100}} \\ \hline
\rowcolor{Gray}
\textbf{Arxiv-MATH} & 0.013 & 6.8 & 0.62 & 90.4 & 0.148 & 26.2 & 0.06 & 25.1 & \textbf{1} & \textbf{100} & 0.11 & 20.8 \\ \hline

\textbf{Arxiv-HEP} & 0.01 & 8.2 & 0.693 & 93.8 & 0.181 & 30.2 & 0.07 & 29.5 & \textbf{1} & \textbf{100} & 0.126 & 23.7 \\ \hline

\rowcolor{Gray}
\textbf{Arxiv-QBIO} & 0.007 & 6.8 & 0.795 & 98.0 & 0.182 & 31.0 & 0.144 & 54.5 & \textbf{1} & \textbf{100} & 0.127 & 24.1 \\ \hline

\textbf{Arxiv-ECON} & 0.01 & 11.4 & 0.772 & 95.9 & 0.196 & 34.8 & 0.142 & 49.8 & \textbf{1} & \textbf{100} & 0.144 & 26.6 \\ \hline

\rowcolor{Gray}
\textbf{Arxiv-CS\_SY} & 0.007 & 5.9 & 0.856 & 99.4 & 0.183 & 31.3 & 0.116 & 47.9 & \textbf{1} & \textbf{100} & 0.132 & 23.9 \\ \hline

\textbf{ICLR} & 0.007 & 5.4 & 0.592 & 91.8 & 0.145 & 29.9 & 0.09 & 41.0 & \textbf{1} & \textbf{100} & 0.124 & 25.1 \\ \hline

\rowcolor{Gray}
\textbf{ACL-ANTH} & 0.002 & 3.1 & 0.74 & 95.1 & 0.123 & 25.2 & 0.034 & 18.2 & \textbf{0.99} & \textbf{100} & 0.096 & 18.2 \\ \hline
\end{tabular}}
\caption{L2 normalization leads to an incremental improvement in performance. Standardization leads to improvement for SciBERT and SPECTER models, but the same effect is not observed for OAG-BERT embeddings.}
\label{apptab:normexp1exp2}
\end{table*}

\section{Analysing Scientific LMs with Textual Neighbors}
\label{sec:app_scilmtneigh}
We present the plots for each of the seven datasets for the experiments with textual neighbors in the following sections.
\subsection{Distribution of Textual Neighbors in the Embedding Space}
\begin{figure*}[htb]
\centering
\setlength{\tabcolsep}{3pt}
\begin{tabular}{ccc} 
    \includegraphics[width=0.32\linewidth]{figures/exp4_1/ACL.png} & \includegraphics[width=0.32\linewidth]{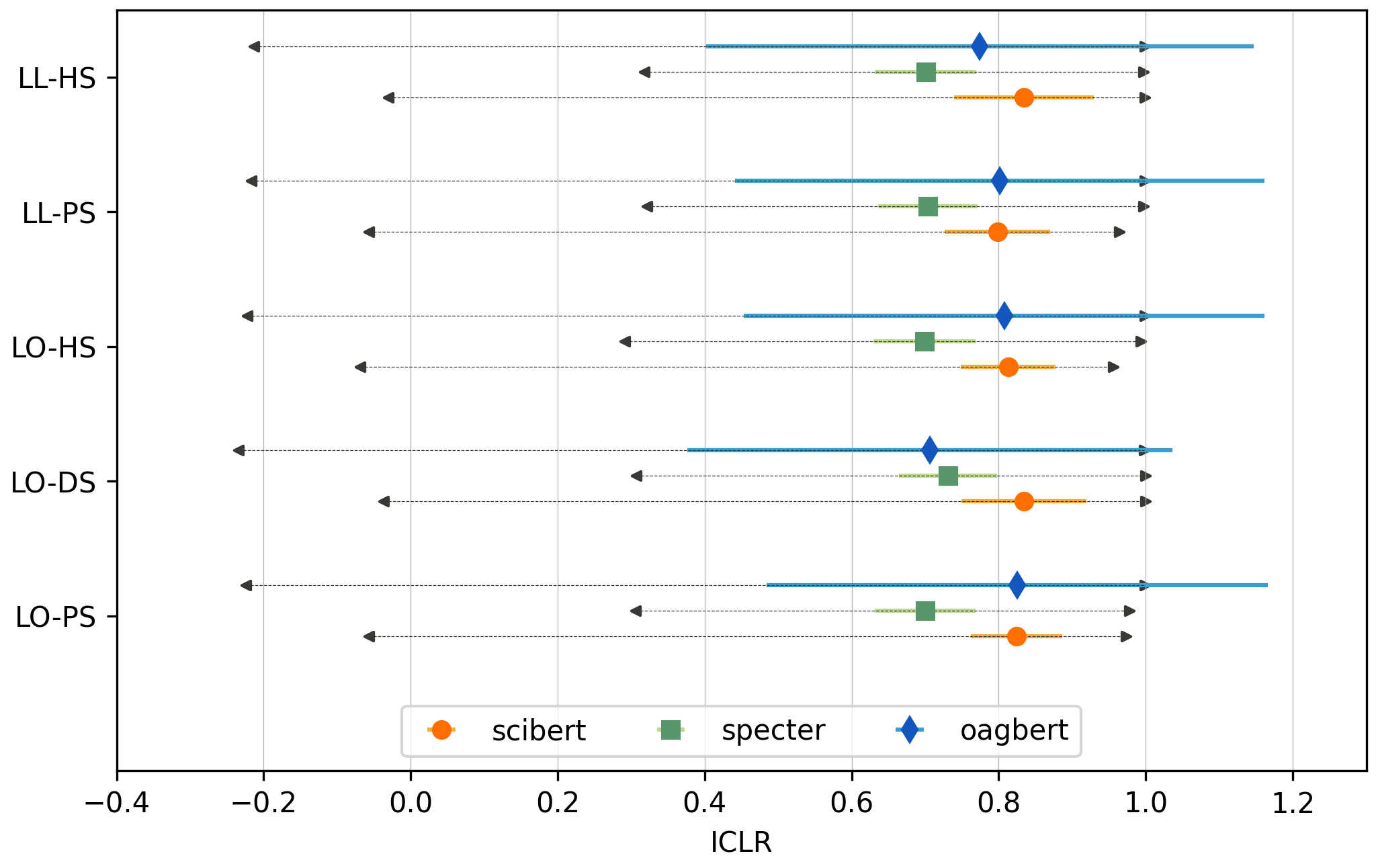} & \includegraphics[width=0.32\linewidth]{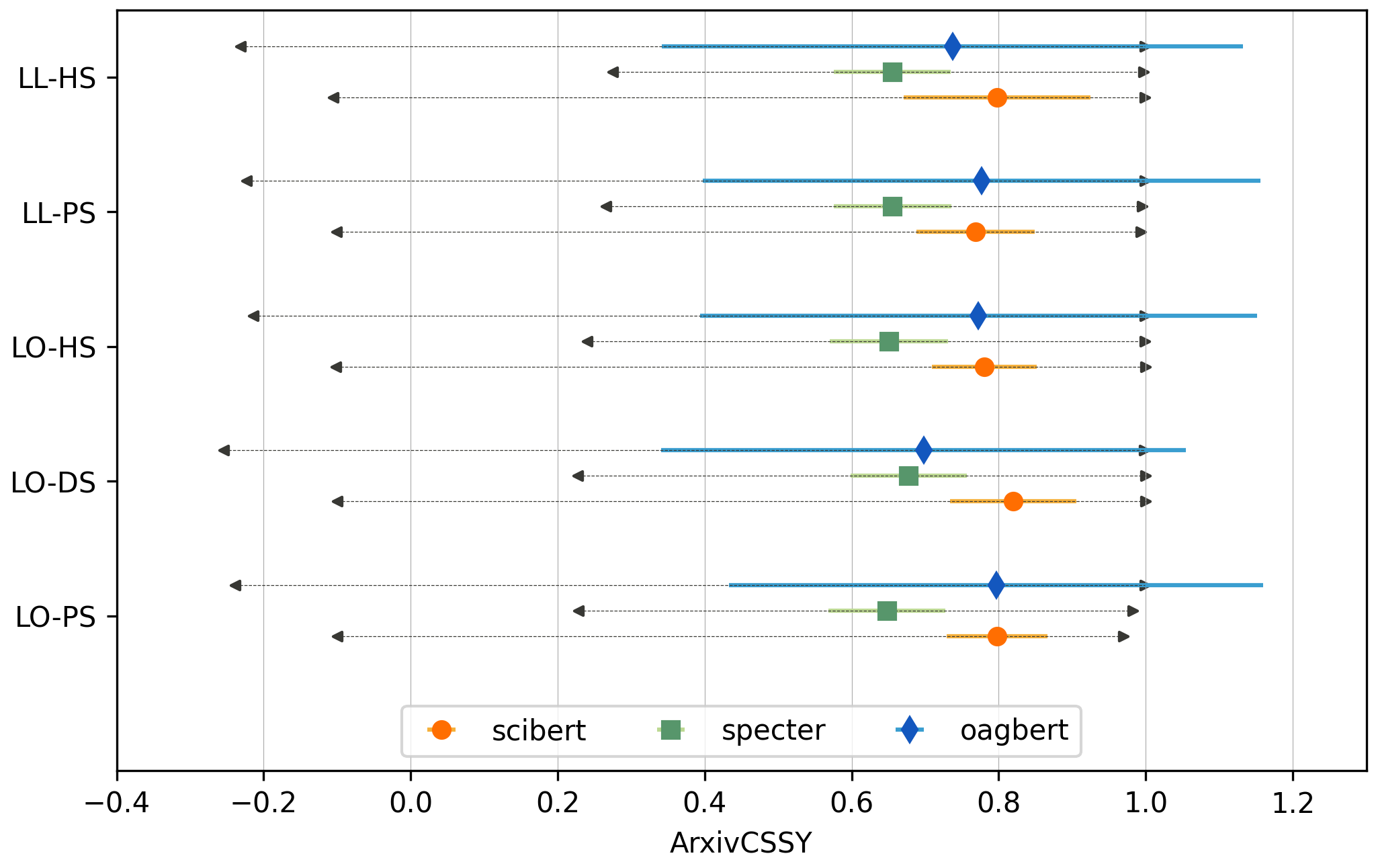} \\
    \includegraphics[width=0.32\linewidth]{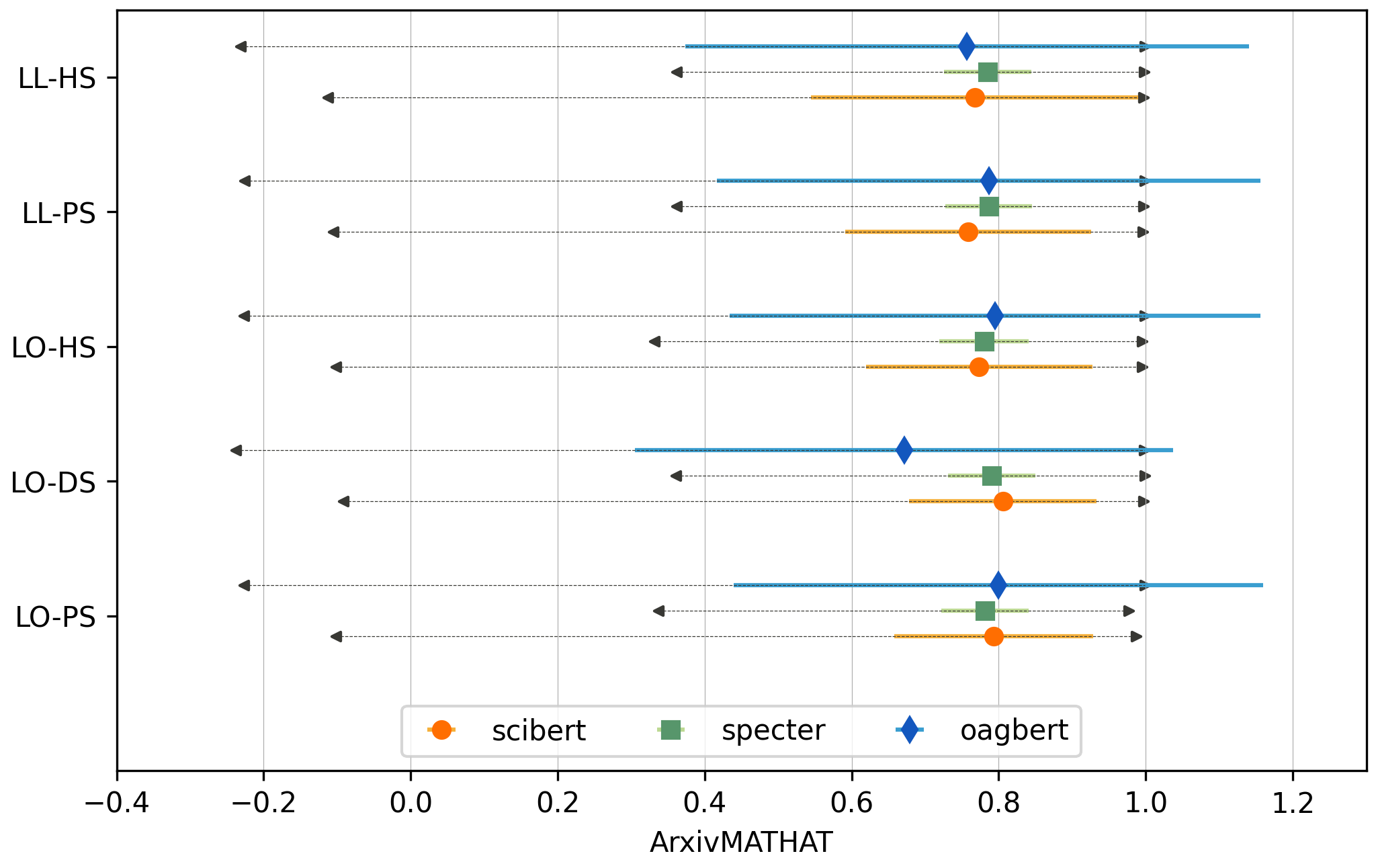} & \includegraphics[width=0.32\linewidth]{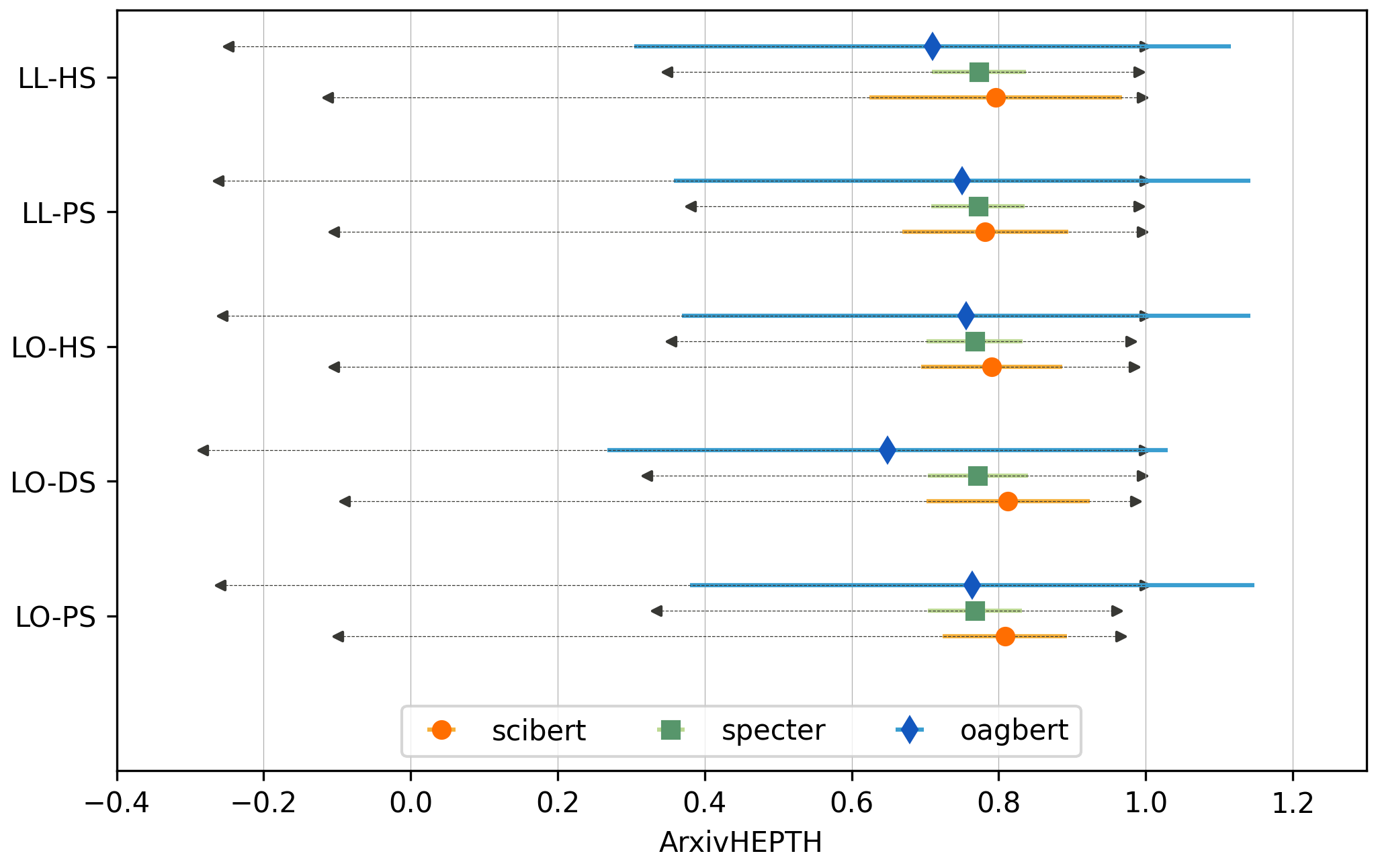} & \includegraphics[width=0.32\linewidth]{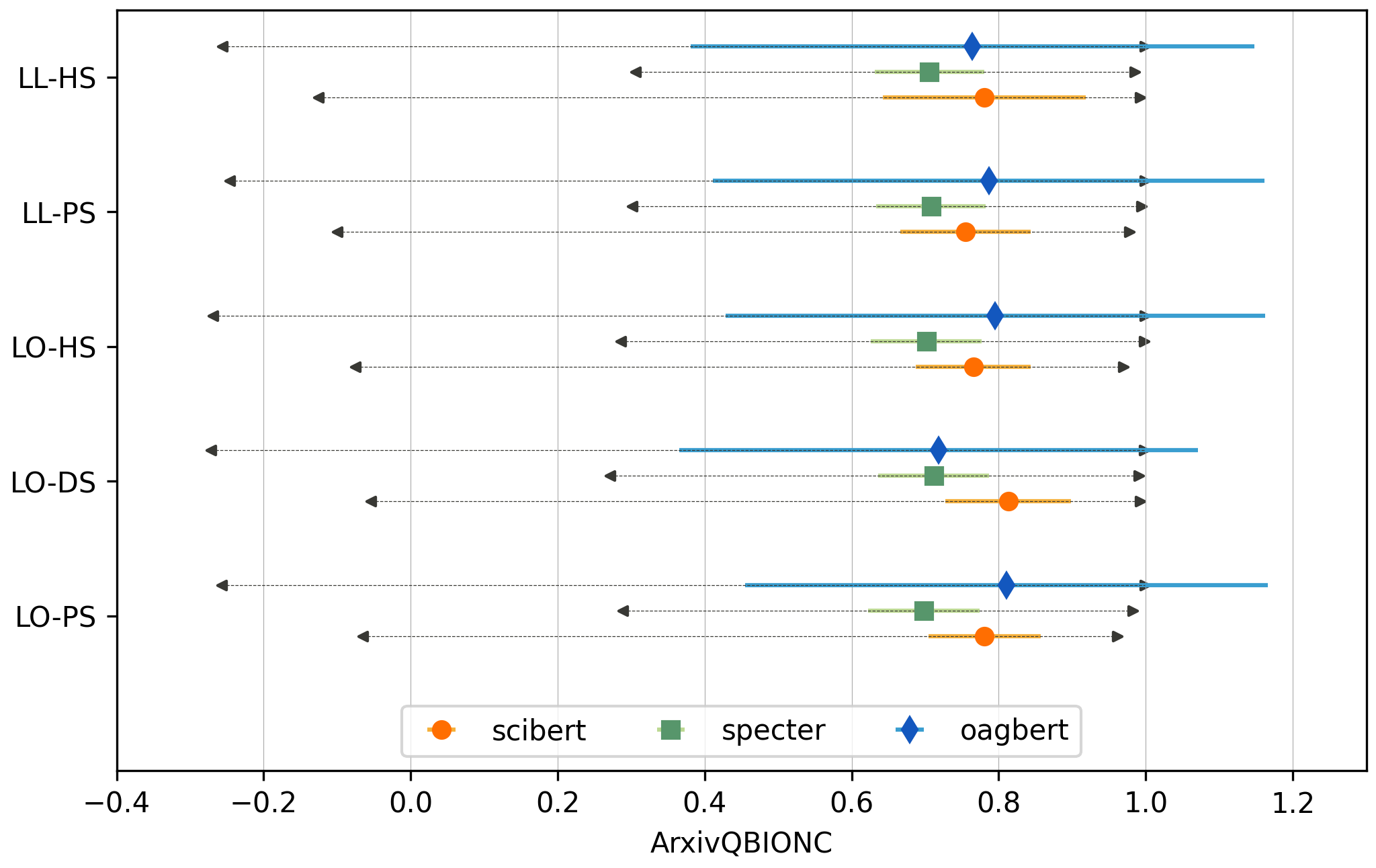} \\
     & \includegraphics[width=0.32\linewidth]{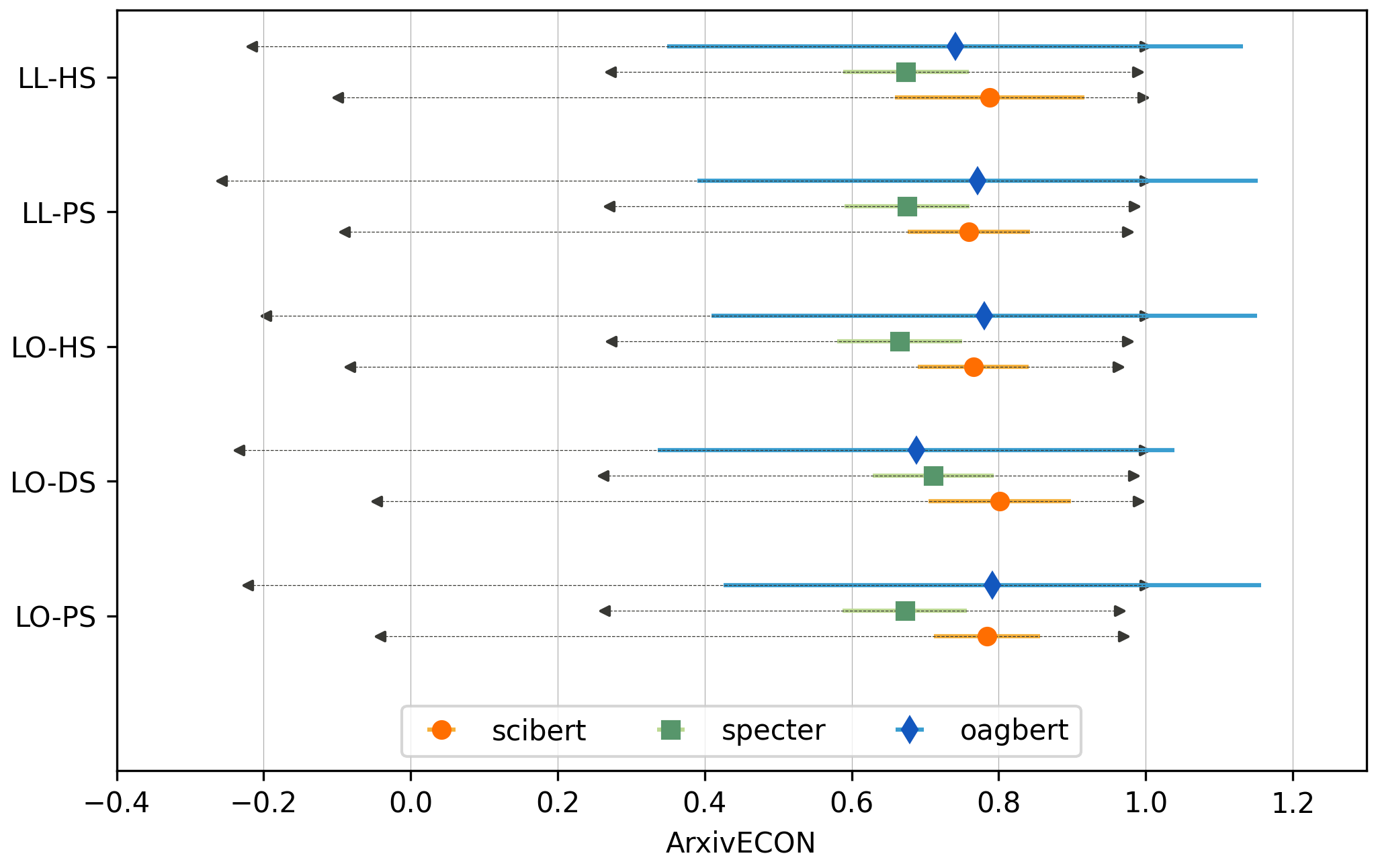} &
\end{tabular}
\caption{Inter Similarity of Textual Neighbor Vectors. The pairwise similarities among documents are spread out in a significantly broader range for OAG-BERT than SciBERT and SPECTER, suggesting that OAG-BERT vectors for different textual neighbors are more spread out in the vector space.}
\label{appfig:geo}
\end{figure*}

\begin{figure*}[htb]
\setlength{\tabcolsep}{2pt}
\centering
\begin{tabular}{ccc} 
    \includegraphics[trim={0.1cm 0.2 0 0},clip,width=0.31\linewidth]{figures/exp4_2/HS_ACL.png} &  \includegraphics[trim={0.1cm 0.2 0 0},clip,width=0.31\linewidth]{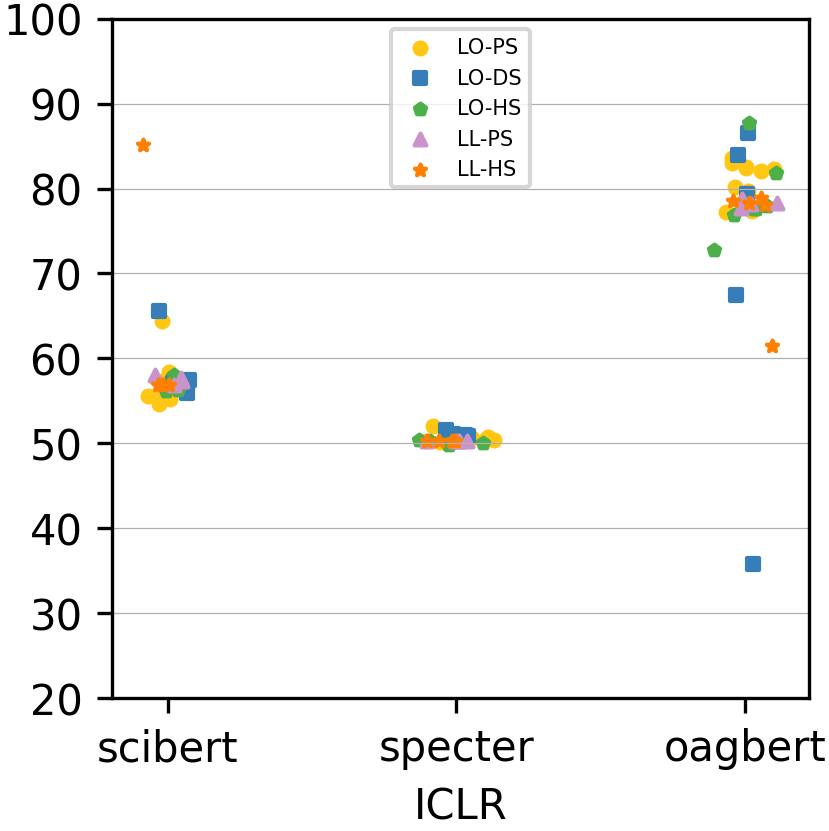} & \includegraphics[trim={0.1cm 0.2 0 0},clip,width=0.31\linewidth]{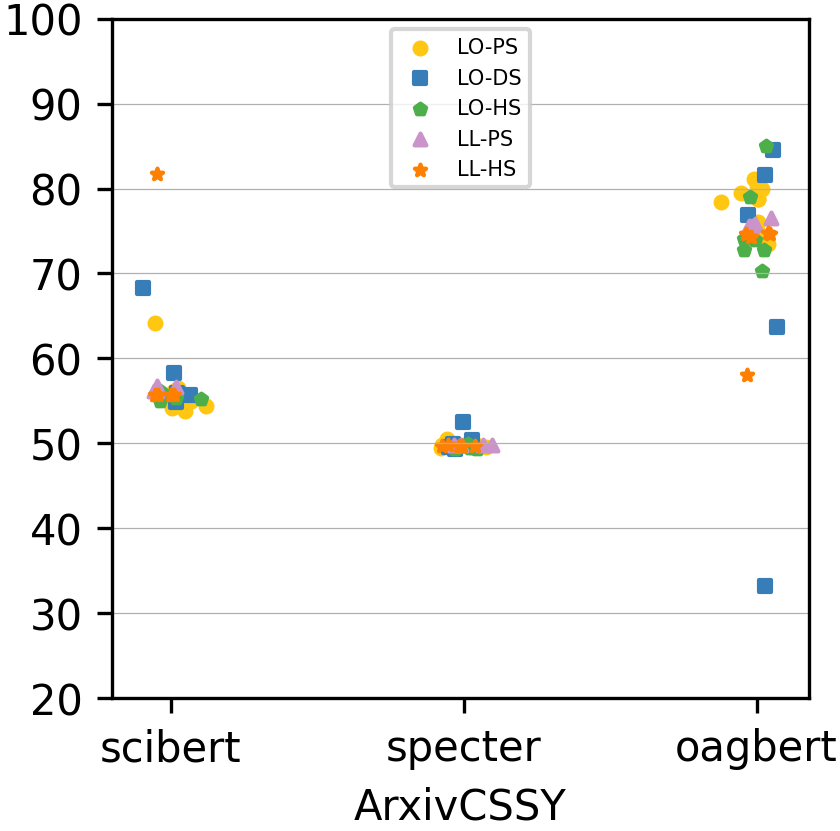} \\
    \includegraphics[trim={0.1cm 0.2 0 0},clip,width=0.31\linewidth]{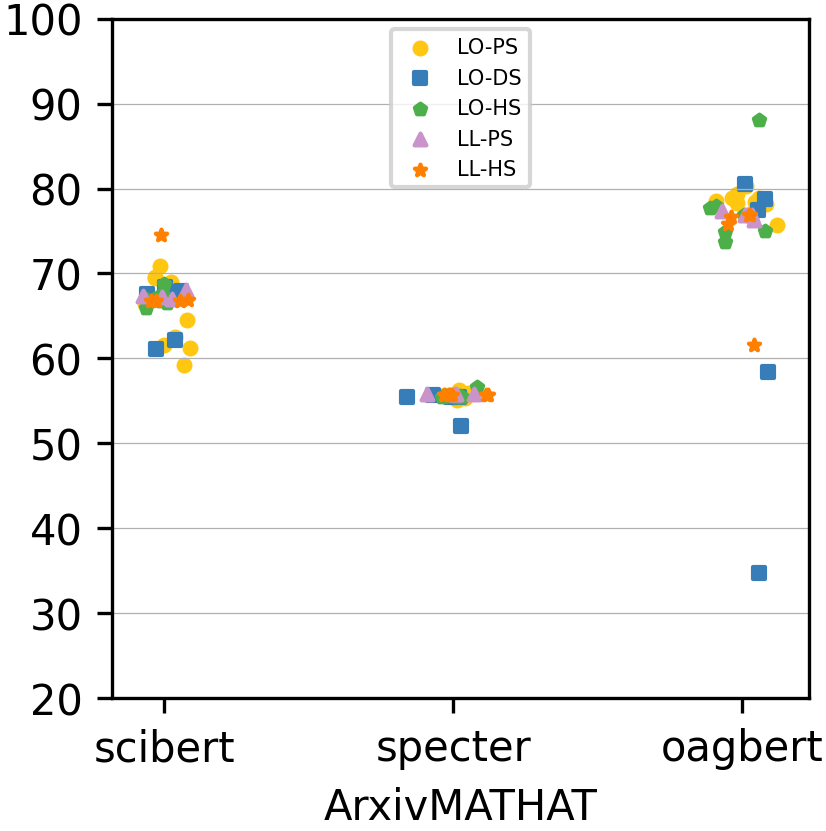} & \includegraphics[trim={0.1cm 0.2 0 0},clip,width=0.31\linewidth]{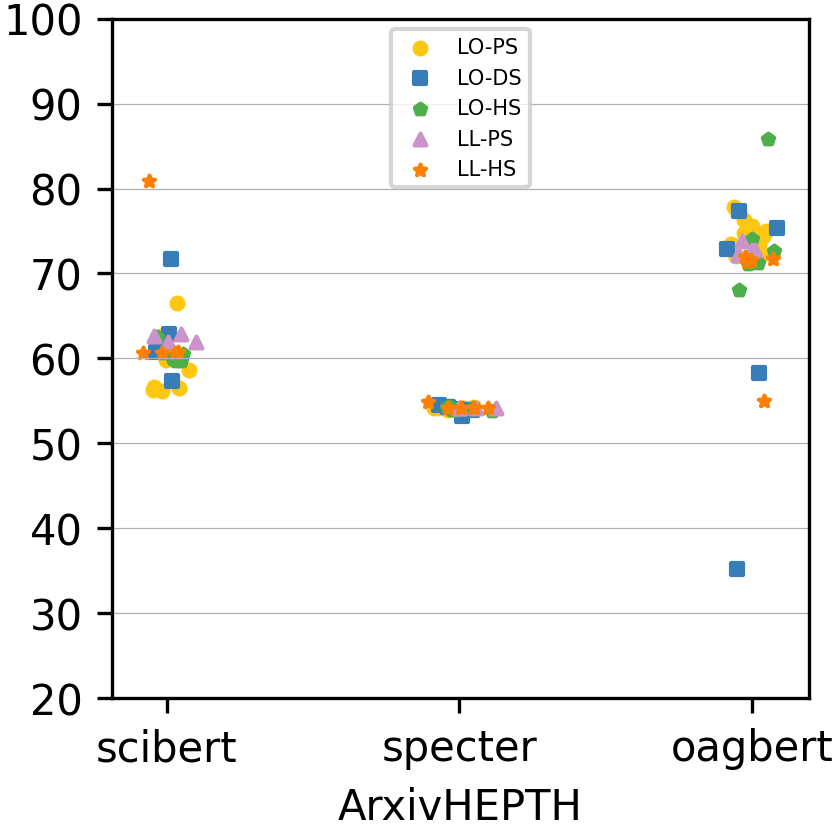} & \includegraphics[trim={0.1cm 0.2 0 0},clip,width=0.31\linewidth]{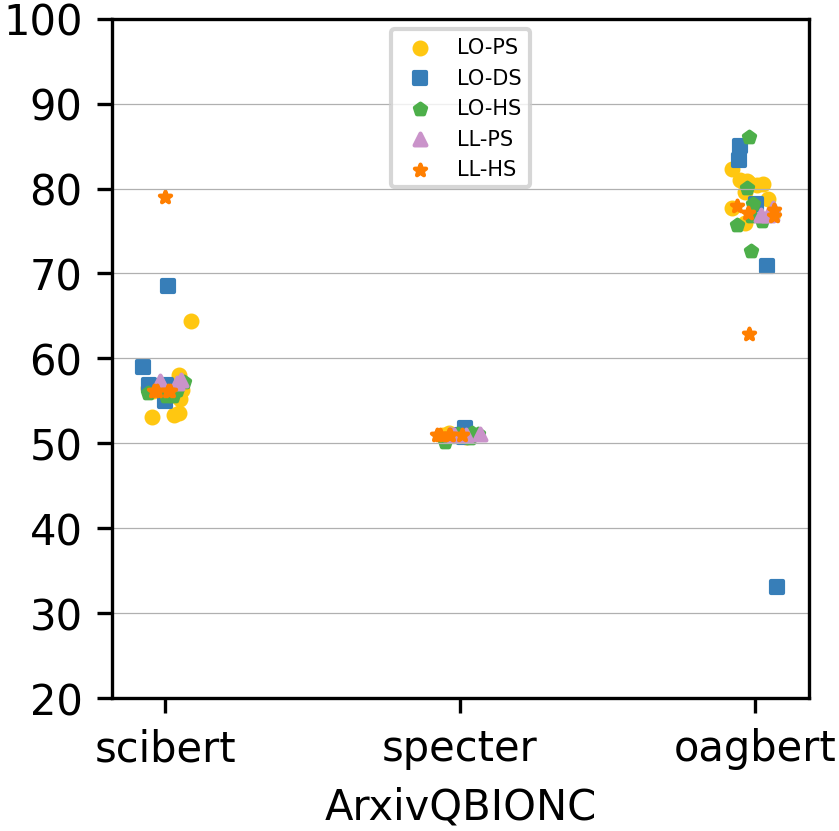} \\
     & \includegraphics[trim={0.1cm 0.2 0 0},clip,width=0.31\linewidth]{figures/exp4_2/HS_ArxivECON.png} &
\end{tabular}
\caption{Percentage of pair of documents for each Textual Neighbor whose similarity is greater than the average similarity. OAG-BERT shows high inter similarity (greater than 50\%) among all textual neighbors suggesting that more than 50\% document pairs have a cosine similarity greater than average similarity.}
\label{appfig:per_hs}
\end{figure*}

We present the plot for inter similarity of textual neighbor vectors in Figure~\ref{appfig:geo}, which depicts the maximum, minimum, mean and standard deviation of all pairs of documents for the five neighbor categories. Next, we present the percentage of document pairs for each of the 32 textual neighbor classes, whose similarity is greater than the average similarity for that particular class in Figure~\ref{appfig:per_hs}.

\subsection{Similarity of Textual Neighbors with Original Documents}
\label{sec:app_simtneighorg}
\begin{figure*}[htb]
\setlength{\tabcolsep}{3pt}
\centering
\begin{tabular}{ccc} 
    \includegraphics[width=0.32\linewidth]{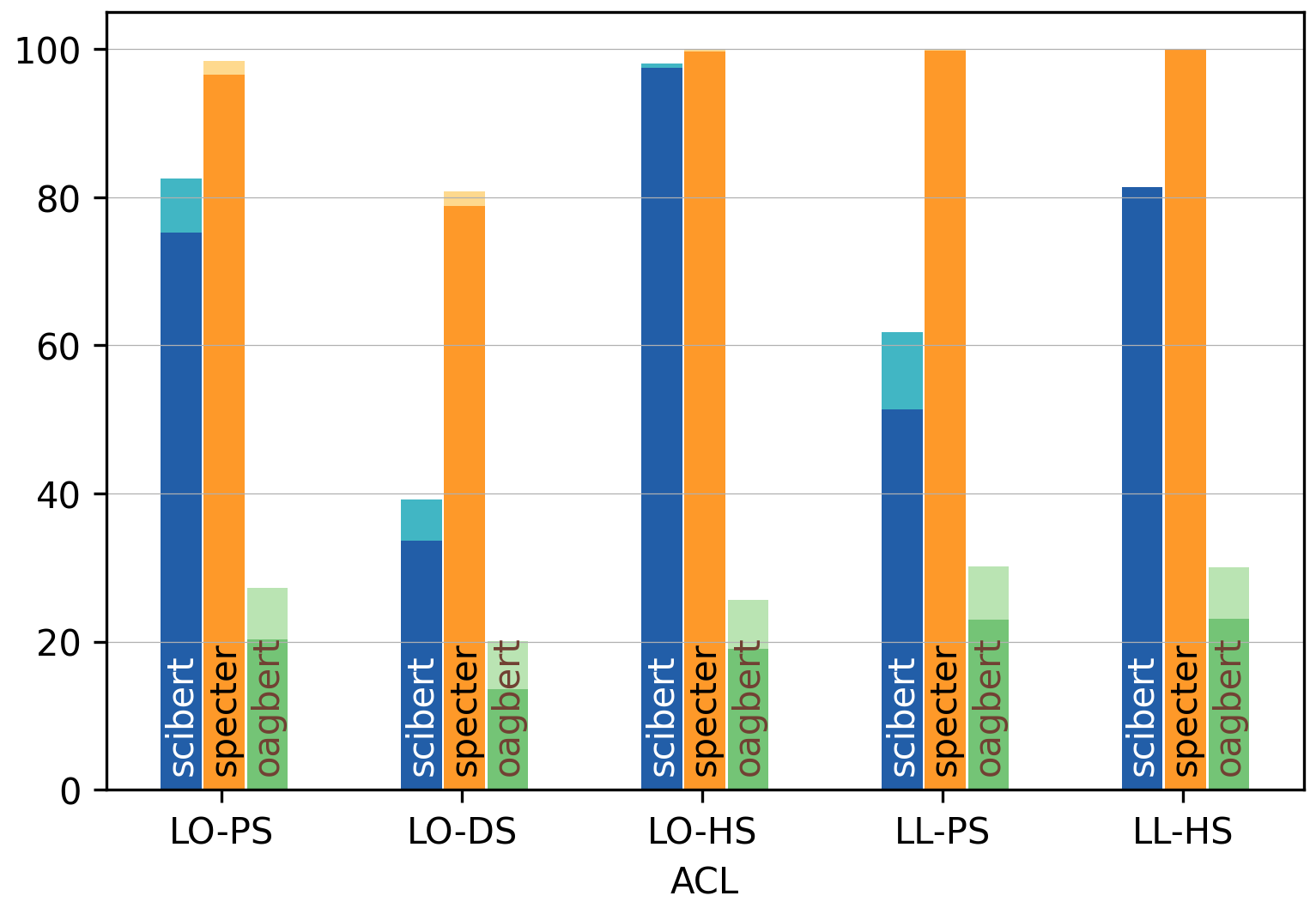} & \includegraphics[width=0.32\linewidth]{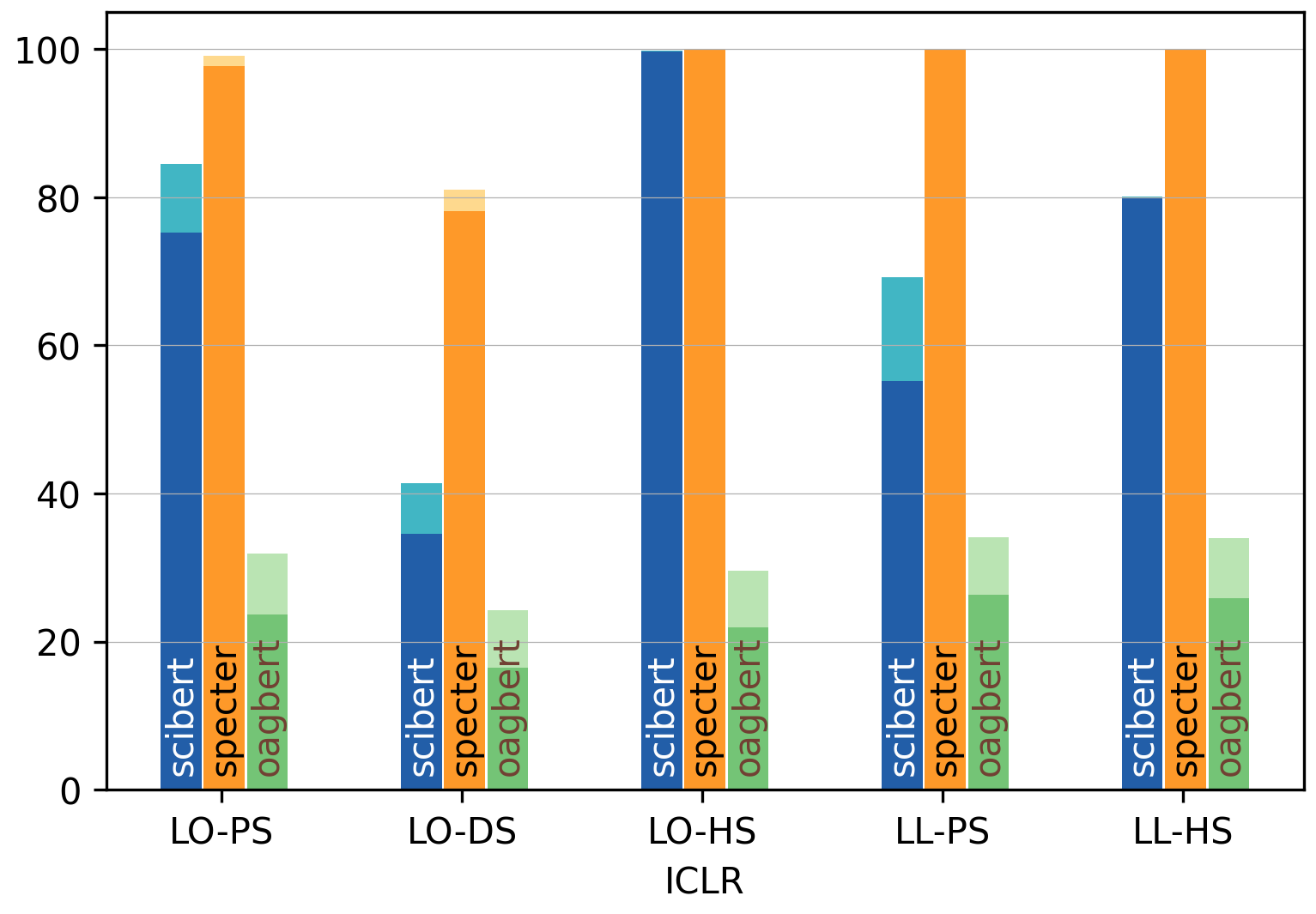} & \includegraphics[width=0.32\linewidth]{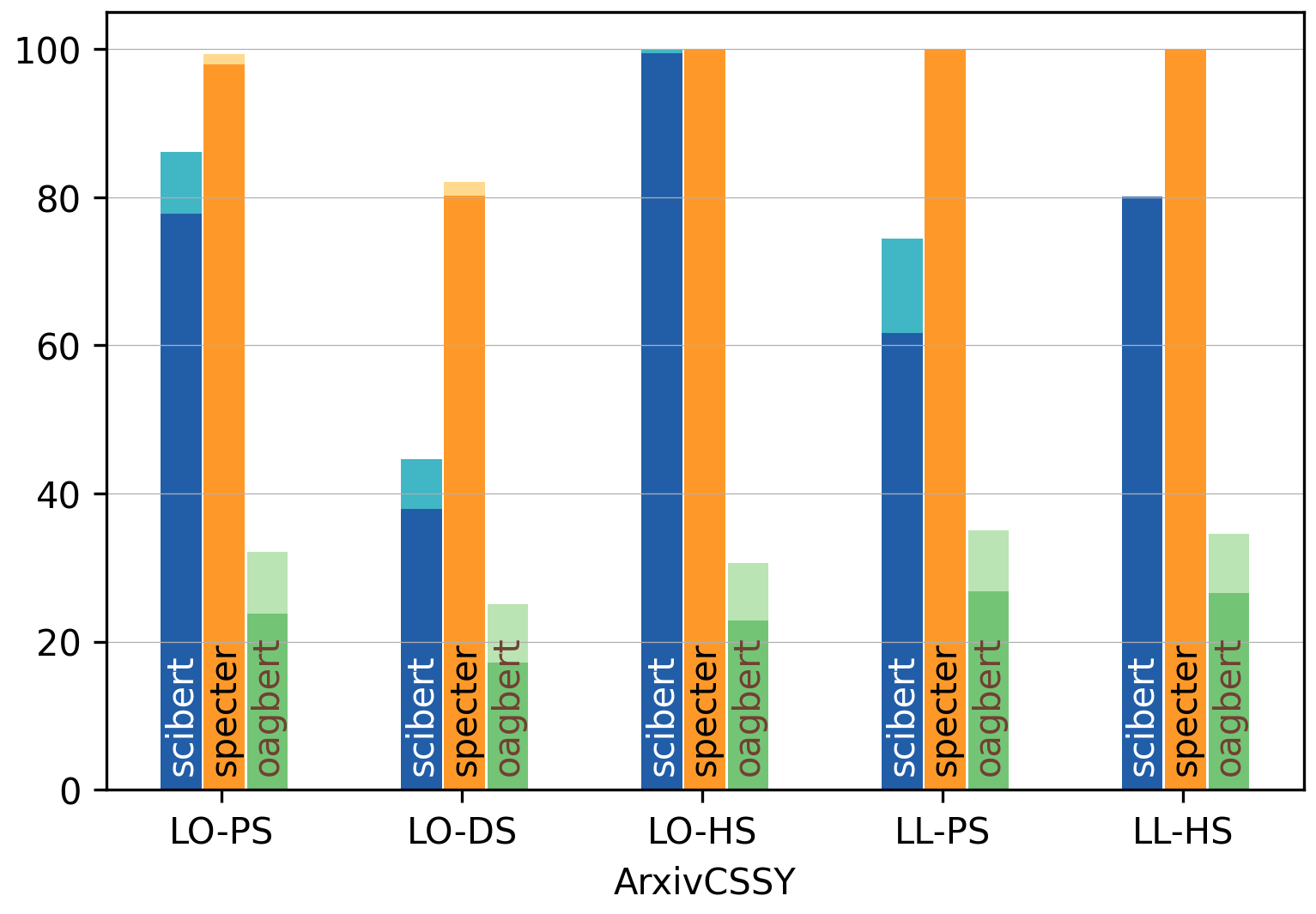} \\
    \includegraphics[width=0.32\linewidth]{figures/exp5/ArxivMATHAT.png} & \includegraphics[width=0.32\linewidth]{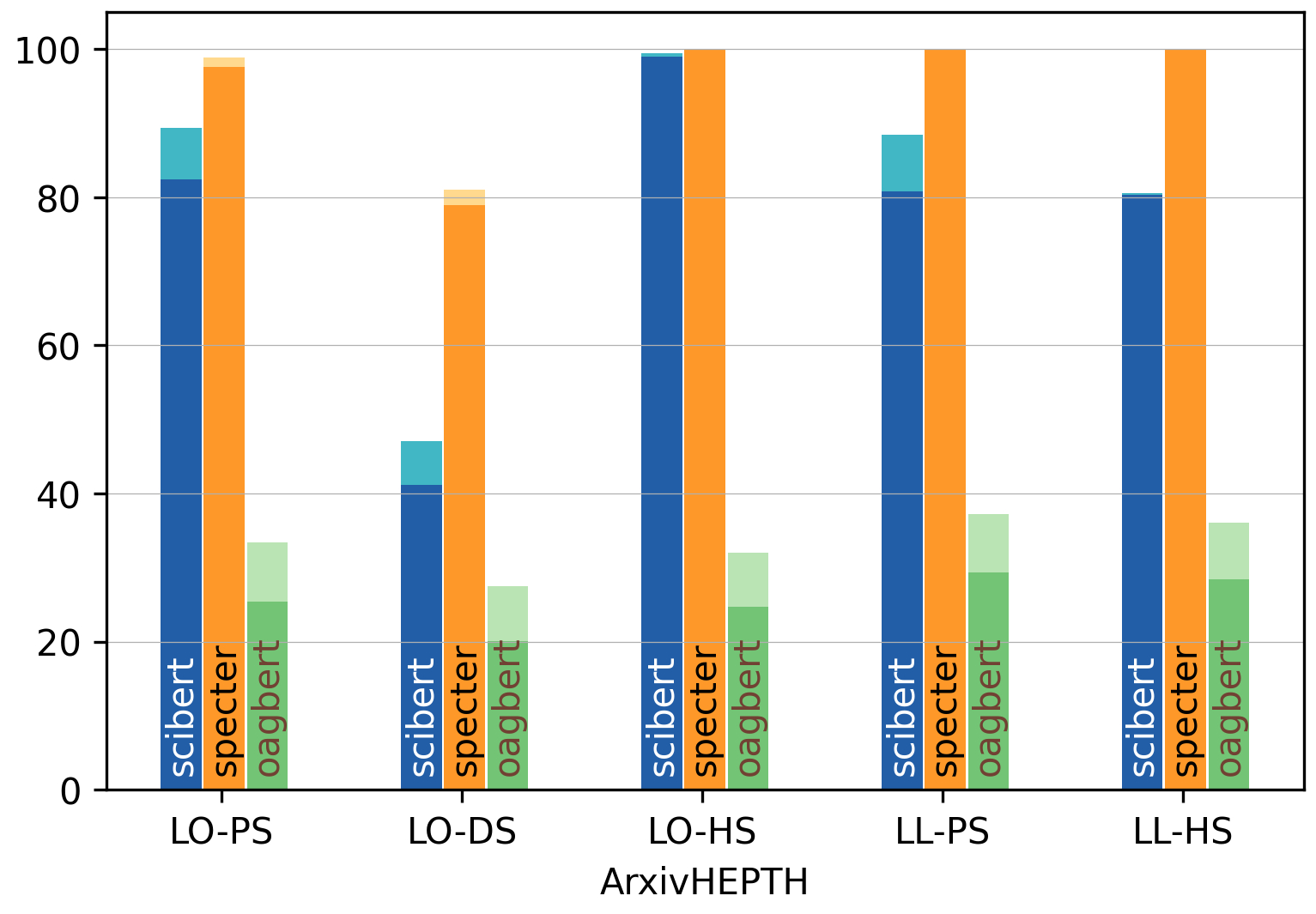} & \includegraphics[width=0.32\linewidth]{figures/exp5/ArxivQBIONC.png} \\
     & \includegraphics[width=0.32\linewidth]{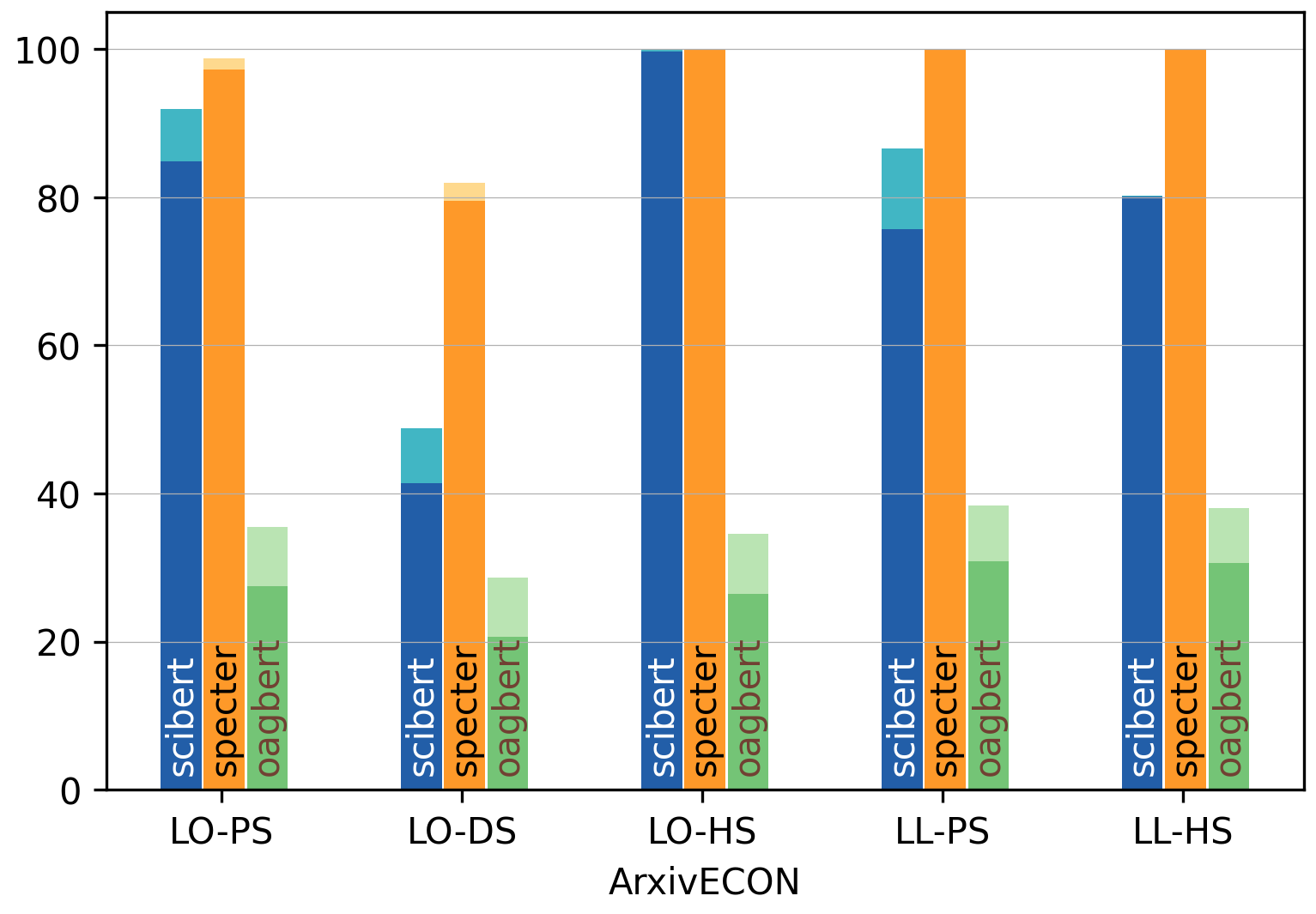} &
\end{tabular}
\caption{Stacked bars representing the percentage of documents of each textual neighbor category which retrieve the original document in the top-1 nearest neighbor and the top-10 nearest neighbor respectively. Results suggest that SciBERT embeddings for textual neighbors of scientific text are the most optimal.}
\label{appfig:exp5}
\end{figure*}

\begin{figure*}[htb]
\setlength{\tabcolsep}{3pt}
\centering
\begin{tabular}{ccc} 
    \includegraphics[width=0.32\linewidth]{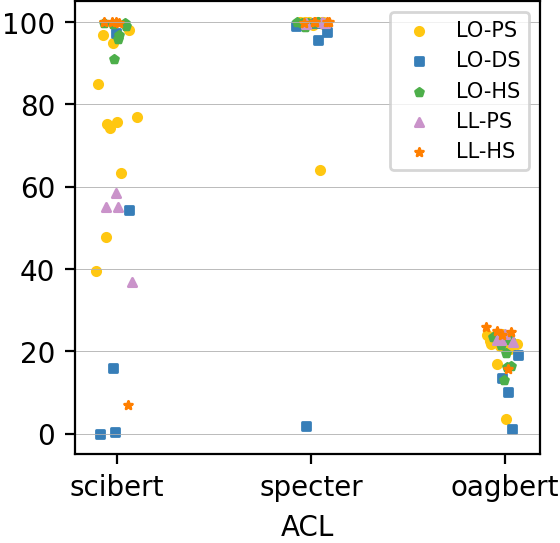} & \includegraphics[width=0.32\linewidth]{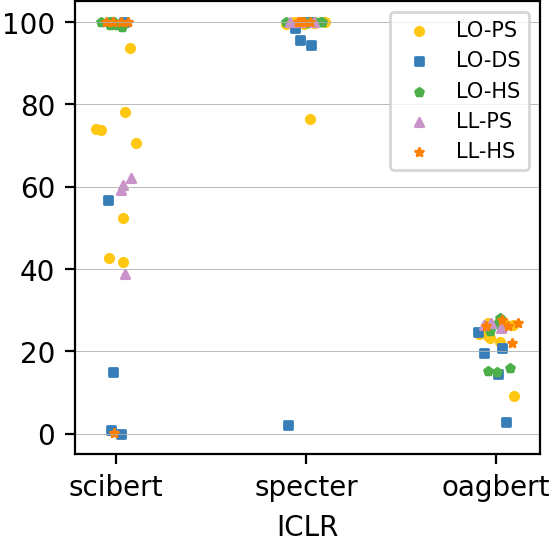} & \includegraphics[width=0.32\linewidth]{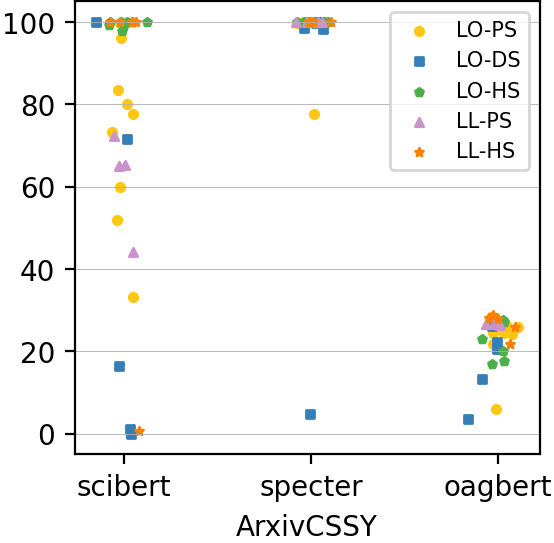} \\
    \includegraphics[width=0.32\linewidth]{figures/exp5_2/ArxivMATHAT.png} & \includegraphics[width=0.32\linewidth]{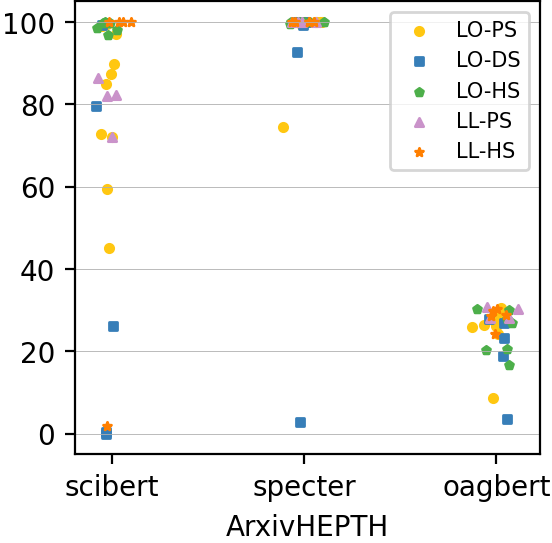} & \includegraphics[width=0.32\linewidth]{figures/exp5_2/ArxivQBIONC.png} \\
     & \includegraphics[width=0.32\linewidth]{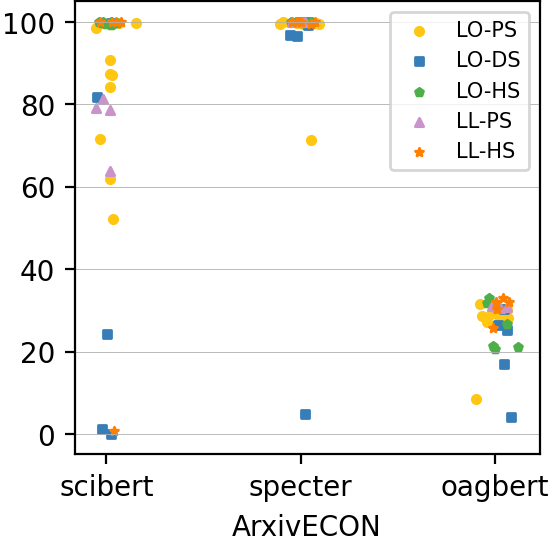} &
\end{tabular}
\caption{Distribution of NN1\_Ret for each textual neighbor category. It can be observed that SciBERT embeddings preserve the hierarchy of NN1\_Ret, i.e. Partially Similar categories (LO-PS and LL-PS) have lower values than Highly Similar categories (LO-HS and LL-HS).}
\label{appfig:exp52}
\end{figure*}

Figure~\ref{appfig:exp5} presents the NN1\_Ret and NN10\_Ret for each of the datasets for the five textual neighbor categories in the stacked bar format (NN10\_Ret stacked onto NN1\_Ret values). It can be observed that for all datasets, SPECTER achieves extremely high ($>$ 95\%) NN1\_Ret values for four (LO-PS, LO-HS, LL-PS, and LL-HS) out of five classes, and hence NN10\_Ret does not significantly improve the retrieval. SciBERT values though less in comparison to SPECTER, show a nice trend where NN10\_Ret does not improve the retrieval significantly for \textit{HS} categories (LO-HS and LL-HS), however it does improve retrieval recall for \textit{PS} categories (LO-PS and LL-PS). OAG-BERT performs poorly with all categories achieving NN10\_Ret values less than 40\%. We present in Figure~\ref{appfig:exp52}, the NN1\_Ret for each of the 32 textual neighbor classes. A close inspection reveals an interesting case for SciBERT, which has extremely low NN1\_Ret values ($<$ 10\%) for one of the LL-HS category, T\_A\_WS whih replaces 50\% whitespace characters randomly with 2-5 whitespaces.

\subsection{Overlap amongst Nearest Neighbors}
\begin{figure*}[htb]
\setlength{\tabcolsep}{3pt}
\centering
\begin{tabular}{ccc} 
    \includegraphics[width=0.32\linewidth]{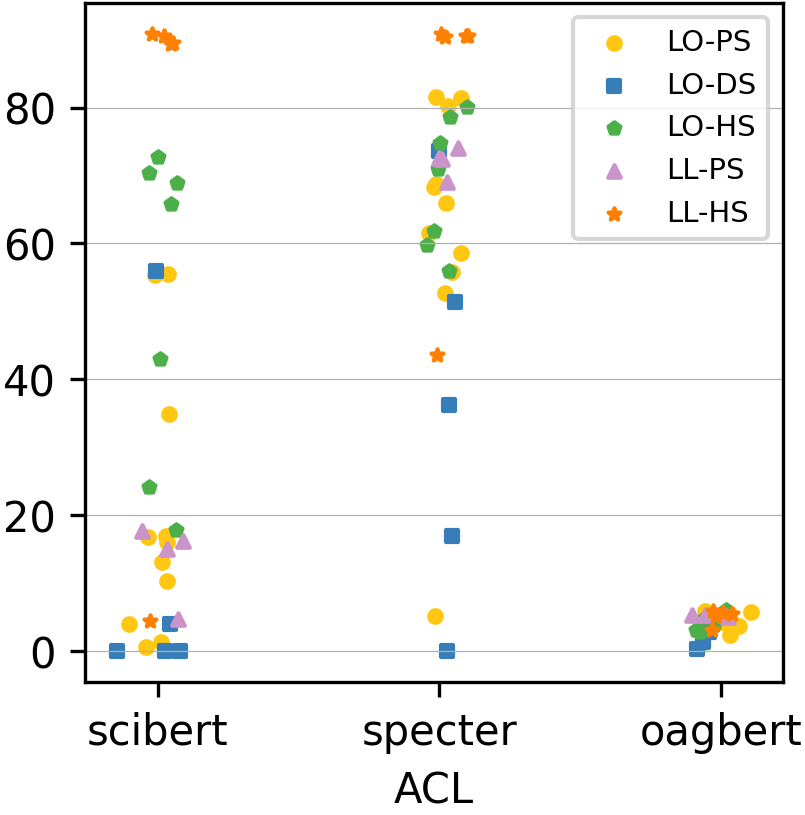} & \includegraphics[width=0.32\linewidth]{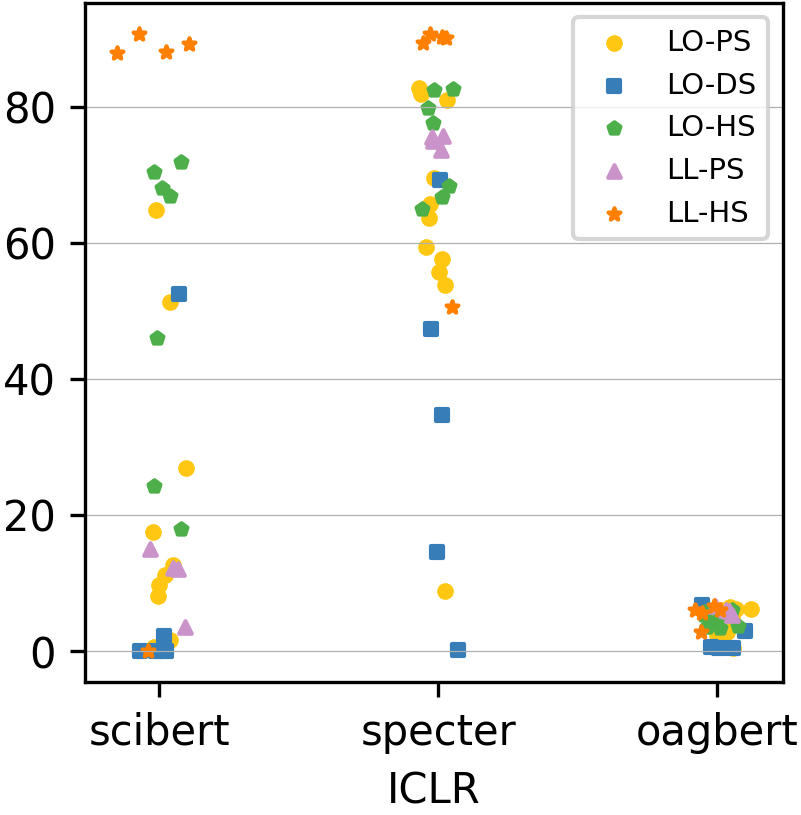} & \includegraphics[width=0.32\linewidth]{figures/exp6/ArxivCSSY.png} \\
    \includegraphics[width=0.32\linewidth]{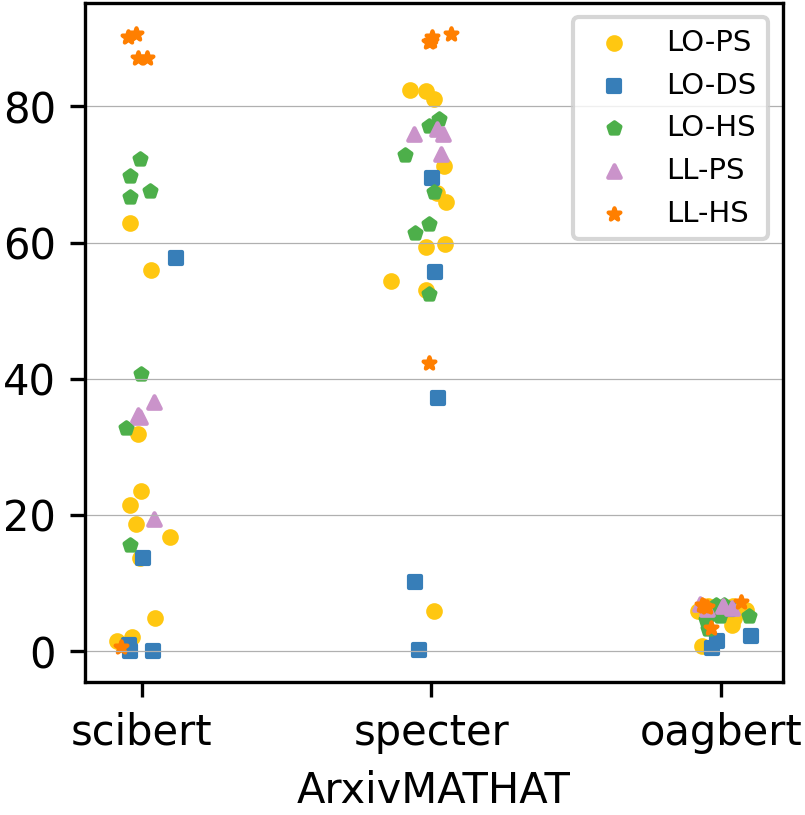} & \includegraphics[width=0.32\linewidth]{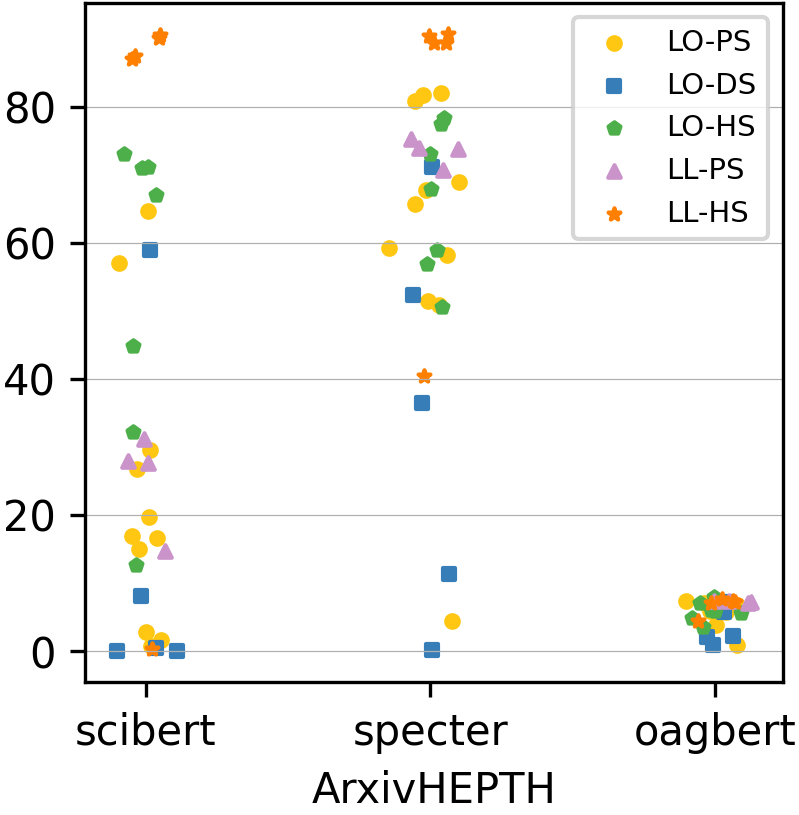} & \includegraphics[width=0.32\linewidth]{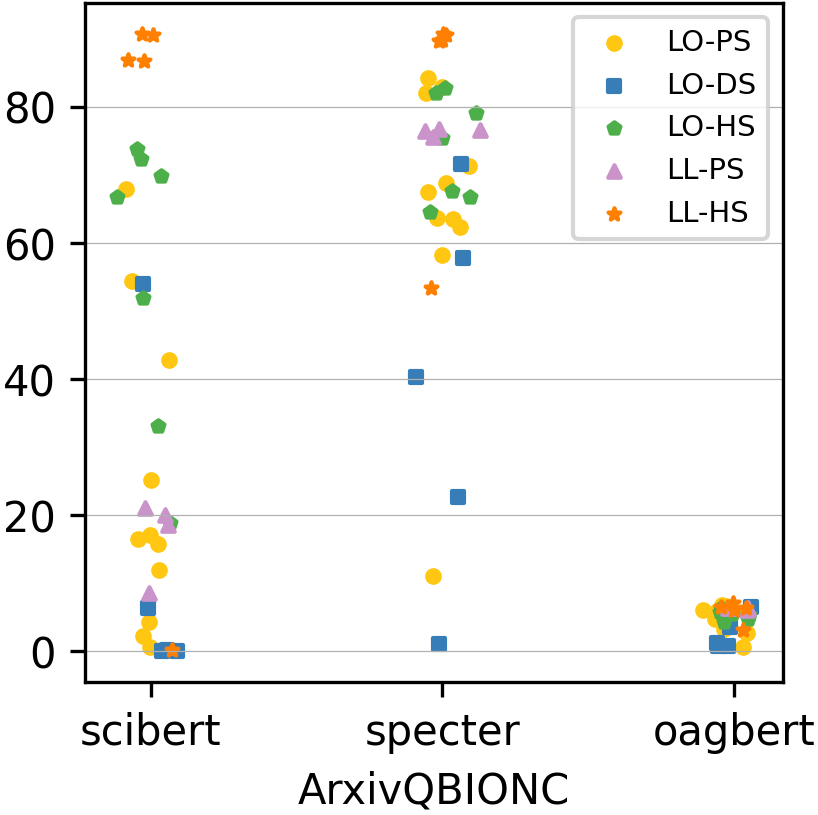} \\
     & \includegraphics[width=0.32\linewidth]{figures/exp6/ArxivECON.png} &
\end{tabular}
\caption{AOP-10 distribution of all categories of textual neighbors. It can be observed that SciBERT performs poorly for the LL-PS category (involves neighbors that scramble sentences such as arranging randomly, or arranging in increasing or decreasing order of sentence length). For the rest of the categories, SciBERT embeddings show desirable order of AOP-10 values, e.g. LL-HS $>$ LO-HS. SPECTER values have high AOP-10, which is desirable, except for the LO-DS category.}
\label{appfig:exp6}
\end{figure*}

AOP-10 is the average overlap percentage among the 10-NN (10 nearest neighbors) of the original document embeddings (T+A) and the textual neighbor embeddings. AOP-10 distribution of the 32 textual neighbor classes is presented in Figure~\ref{appfig:exp6}. Low overlap percentage for OAG-BERT suggests that the model falters when presented with textual neighbors and does not represent textual neighbors in the neighborhood of the original document embeddings in the embedding space.

\end{document}